\documentclass[12pt]{extarticle}
\usepackage[T1]{fontenc}
\usepackage[utf8]{inputenc}
\usepackage{graphicx}
\usepackage{float}

\usepackage[margin=1.2in]{geometry}

\usepackage[font={footnotesize}, margin=0.75cm, labelfont=bf]{caption}

\usepackage[colorlinks]{hyperref}
\hypersetup{linkcolor=black,citecolor=black,filecolor=cyan,urlcolor=black}

\usepackage{natbib}
\usepackage{nameref}

\newcommand{\chla}{Chl~\textit{a}}
\newcommand{\changed}[1]{#1}

\title{The peak absorbance wavelength of photosynthetic pigments
around other stars from spectral optimization}
\author{
    \parbox{\linewidth}{\centering
        Owen R. Lehmer, David C. Catling, Mary N. Parenteau, Nancy Y. Kiang, and Tori M. Hoehler
    }
}

\date{Front. Astron. Space Sci., 08 July 2021, https://doi.org/10.3389/fspas.2021.689441}

\begin{document}

\maketitle

\begin{abstract}
In the search for life on other planets, the presence of photosynthetic surface
vegetation may be detectable from the colors of light it reflects, which on the
modern Earth is a steep increase in reflectance between the red and
near-infrared wavelengths, a ``red edge.'' This edge-like signature
occurs at wavelengths of peak photon absorbance, which are the result of
adaptations of the phototroph to their spectral environment. On planets
orbiting different stellar types, red edge analogs may occur at other
colors than red. Thus, knowing the wavelengths at which photosynthetic
organisms preferentially absorb and reflect photons is necessary to detect
red edge analogs on other planets. Using a numerical model that predicts
the absorbance spectrum of extant photosynthetic pigments on Earth from
\citet{marosvolgyi_cost_2010}, we calculate the absorbance spectrum for
pigments on an Earth-like planet around F through late M type stars that are
adapted for maximal energy production. In this model, cellular energy
production is maximized when pigments are tuned to absorb at the wavelength
that maximizes energy input from incident photons while minimizing energy
losses due to thermal emission and building cellular photosynthetic apparatus.
We find that peak photon absorption for photosynthetic organisms around F type
stars tends to be in the blue while for G, K, and early M type stars, red or
just beyond is preferred. Around the coolest M type stars, these organisms may
preferentially absorb in the near-infrared, possibly past 1 micron. These
predictions are consistent with previous, qualitative estimates of pigment
absorptance. Our predicted absorbance spectra for photosynthetic surface
organisms depend on both the stellar type and planetary atmospheric
composition, especially atmospheric water vapor concentrations, which alter the
availability of surface photons and thus the predicted pigment absorption. By
constraining the absorbance spectra of alien, photosynthetic organisms, future
observations may be better equipped to detect the weak spectral signal of red
edge analogs.
\end{abstract}

\section{Introduction}\label{sec:introduction}
Oxygenic photosynthetic organisms dominate primary productivity on the modern
Earth \cite[e.g.][]{field_primary_1998}. Abundant, green surface vegetation is
a prominent example of such productivity. While surface vegetation slightly
reflects green photons in the visible spectrum, the strongest spectral
reflectance occurs just beyond the visible in  near-infrared (NIR) wavelengths.
This sharp contrast in absorbance in the red and reflectance in the NIR is
termed the ``vegetation red edge'' or simply the ``red edge.'' This spectral
signature is sufficiently strong on the modern Earth and distinct from mineral
spectra that satellite observations of it are used to map the presence and
activity of vegetation \citep{tucker_sensor_1976}.  It thus serves as a model
for  similar reflectance biosignatures that future telescope observations could
detect on an Earth-like exoplanet \citep[e.g.][]{sagan_search_1993,
seager_vegetations_2005, kiang_spectral_2007, omalley-james_vegetation_2018,
wang_diurnal_2021}.

The absorption of red photons by green land plants is due to chlorophyll
\textit{a} (\chla{}), which is the primary photopigment in oxygen-producing
photosynthetic organisms, such as algae and cyanobacteria on the modern
Earth.  In these organisms, \chla{} and other light harvesting pigments absorb
photons across the visible, with the long wavelength cut-off where \chla{}
absorbs red photons, peaking at 680 and 700 nm.  In plant leaves, strong
scattering occurs in the NIR due to the change in the index of
refraction between leaf mesophyll cell walls and air spaces in the leaf and
lack of absorbance by pigments \citep{gausman_leaf_1974}. This step-like
change in spectral absorbance results in the red edge.  

In all phototrophs, accessory pigments often accompany the primary
photopigment, providing additional photon harvesting. Photon absorption for
\chla{} peaks in the blue and red, and accessory pigments offer additional
spectral coverage. The excitation energy of shorter wavelength, higher-energy
photons is stepped down through energetically favorable resonant transfer to
the red absorbance of \chla{}, which trap that energy in a reaction center
\citep[e.g.][]{blankenship_basic_2008}. The excess photon energy is released as
heat during this transfer process so that each photon absorbed by the
photosynthetic cell provides an input energy to \chla{} equivalent to a red
photon. The reaction center \chla{} molecules enable charge separation; that
is, the excited pigment provides the conversion of photon energy to chemical
potential, giving up an excited electron for electron transfers in biochemical
reactions. Around other stars, \chla{} analogs could arise in the form of
different pigments that may be tuned to absorb photons at different
wavelengths, changing the wavelength of a vegetative red edge analog.  See
\citet{schwieterman_exoplanet_2018} and \citet{blankenship_basic_2008,
blankenship_molecular_2014} for thorough discussions of \chla{} and the
molecular mechanisms of photosynthesis.

The structure of \chla{} has been attributed to evolution from simpler
porphyrins, or the chemical properties required of a pigment to participate in
oxygenic photosynthesis \citep[e.g.][]{mauzerall_chlorophyll_1976,
bjorn_viewpoint:_2009,bjorn_spectral_2015}. In addition, optimization arguments
based on the available radiation spectrum have been proposed as a driver for
the structure and absorbance spectrum of \chla{} \citep{bjorn_why_1976,
stomp_colorful_2007, kiang_spectral_2007-1, milo_what_2009,
marosvolgyi_cost_2010, ritchie_could_2017, arp_quieting_2020}.
\citet{kiang_spectral_2007}, \citet{stomp_colorful_2007}, and
\citet{ritchie_could_2017} observed that the absorption wavelength for reaction
center pigments in extant organisms correlates with spectral peaks in the
incident photon flux per uniform wavelength interval. However,
\citet{bjorn_why_1976} and \citet{milo_what_2009} noted that such correlations
may be somewhat artificial, as the wavelength of peak photon flux depends on
whether one considers photons per uniform frequency interval, or photons per
uniform wavelength interval. Therefore, optimality arguments based on maximal
rates of energy storage have been explored \citep{bjorn_why_1976,
milo_what_2009, marosvolgyi_cost_2010}.

\citet{bjorn_why_1976} considered how incident flux constrains the maximum
achievable chemical potential in optically thin systems, such as in unicellular
photosynthetic organisms. By approximating the Sun as a blackbody and
accounting for thermal emission of pigments, \citet{bjorn_why_1976} showed
that, in full sunlight, chemical energy production is maximized in
photosynthetic surface organisms when absorbing near 700 nm. A similar optimal
wavelength was found by \citet{milo_what_2009} by considering the energetic
overhead encountered during photosynthesis to move electrons from H$_2$O to
NADPH, termed an ``overpotential.'' Using the empirical overpotential of
Photosystems I and II along with the spectrally resolved flux available at the
Earth's surface, \citet{milo_what_2009} found pigment absorption was optimized
between 680 and 720 nm.  \citet{marosvolgyi_cost_2010} extended the work of
\citet{bjorn_why_1976} to consider optically thick systems, accounting for the
possibility of multiple dipoles, and the spectrally resolved incident Solar
flux considered by \citet{milo_what_2009}. In addition,
\citet{marosvolgyi_cost_2010} considered the energetic cost for a cell to
create pigments and store photosynthetic energy, an important consideration
raised by \citet{milo_what_2009}. With these enhancements to the model of
\citet{bjorn_why_1976}, \citet{marosvolgyi_cost_2010} showed that \chla{} is
optimized when peak absorption occurs at $\sim 680$ nm, the $Q_y$ band of
\textit{in vivo} \chla{}, and that the measured absorption profile of \chla{}
around that peak is similarly optimal. In addition, substituting the surface
photon flux for the ambient photon flux in the water column,
\citet{marosvolgyi_cost_2010} found that the same pigment optimization model
reproduced the absorbance profile of \textit{Rhodobacter sphaeroides}
chromatophores in the NIR.

While \citet{marosvolgyi_cost_2010} optimized power gain based on incident flux
to predict pigment absorbance spectra, \citet{arp_quieting_2020} assumed a
fixed power demand by photosynthetic organisms to determine the optimal
wavelength for pigment absorption. \citet{arp_quieting_2020} examined the role
of antenna light harvesting pigments in picking out separate spectral bands of
an organism's light environment to minimize the mismatch between the energy
supply and demand. A two-absorber system at separate bands helps to ensure a
steadier power supply, which can be generalized to more absorbers.  This scheme
appears to predict wavelengths of peak absorbance for both the trap and Soret
band wavelengths for \chla{}-bearing organisms, but the role of light quality
for underwater and anoxygenic phototrophs is less clear.  Because this scheme
does not directly model the actual light harvesting processes and requires
assumptions of the shape of the pigment absorbance peaks and the number of
absorbers, we adopt the model of \citet{marosvolgyi_cost_2010} to be agnostic
to these specifications and to more directly simulate light energy use
processes in this work. The model of \citet{marosvolgyi_cost_2010} requires
only the available photon flux to predict plausible pigment absorbance spectra,
thus it could constrain the absorbance spectra of surface pigments around other
stars.

Prediction of pigment absorption around other stars has been considered
previously, but via a simplified model with stars approximated as blackbodies
\citep{lehmer_productivity_2018}, or qualitatively based on the incident
spectral flux \citep{kiang_spectral_2007, ritchie_could_2017,
takizawa_red-edge_2017}. Based on the observed correlations between photon flux
and pigment peak absorbance wavelengths, \citet{tinetti_detectability_2006}
proposed a red edge analog for an M dwarf star could be near 1.1 $\mu$m.
\citet{takizawa_red-edge_2017} also explored photosynthesis around M dwarf stars,
but considered how multi-band photosystems might be restricted in different
spectral light environments and did not attempt to predict light harvesting
pigment spectra precisely.

Here, we apply the power gain optimization model of
\citet{marosvolgyi_cost_2010} to the Earth around other stars to predict the
absorbance spectra of alien photoautotrophs. Using simulated surface spectra
through the modern Earth's atmosphere for F2V through M5V stars from
\citet{kiang_spectral_2007}, we predict the absorption peak and profile of
pigments for various stellar types. Our predictions could inform future
observations looking for the photosynthetic red edge as a possible
biosignature \citep{sagan_search_1993, seager_vegetations_2005,
kiang_spectral_2007, omalley-james_vegetation_2018, wang_diurnal_2021} and
provide a quantitative update to the qualitative predictions for red edge
analogs around other stars from \citet{kiang_spectral_2007}.

\section{Methods}\label{sec:methods}

To predict optimal pigment absorption based on incident photon flux, we use the
model initially developed by \citet{bjorn_why_1976} then refined and justified
in detail by \citet{marosvolgyi_cost_2010}. Our implementation recreates the
numerical model described in \citet{marosvolgyi_cost_2010}, which we summarize
below. 

The model solves for an absorbance spectrum that optimizes power gain in an
optically thick suspension of photosynthetic cells. It does so by considering
the incident flux available for absorption, thermal emission, and the costs to
build the photosynthetic apparatus. With incident flux known, the input energy
intensity over all frequencies in W m$^{-2}$, $P_{in}$, is given by

\begin{equation}\label{eqn:pin}
    P_{in}=J_{L}\cdot\mu=J_{L}\cdot kT \cdot \ln{\left( \frac{J_{L}}{J_{D}}\right)}
\end{equation}
where $k$ is the Boltzmann constant ($1.381\times10^{-23}$ J K$^{-1}$), $T$ in
K is the temperature of the organism (assumed to be at room temperature, 295
K), $J_{L}$ is the excitation rate in light (s$^{-1}$ m$^{-2}$), and $J_{D}$ is
the thermal excitation rate (s$^{-1}$ m$^{-2}$)
\citep{ross_thermodynamics_1967, marosvolgyi_cost_2010}. In equation
\ref{eqn:pin}, $\mu$, in Joules, is the potential energy of the excited
pigments in the system under external illumination. Thus, $P_{in}$ is the
number of excitation events (photon absorption) generating an excited pigment
multiplied by the potential chemical energy of such a pigment, $\mu$. 

The excitation rate due to the incident stellar flux, $J_{L}$, is found by
dividing the incident flux into $n$ equally sized frequency bins then
multiplying by the pigment absorptance for the given frequency. The bin size
should be small enough such that spectral features are not obscured. However,
large discontinuities between adjacent bins can hinder convergence, which can
occur if too many bins are used. Dividing the spectra in this work, which
includes frequencies between ${\sim}$1200 and ${\sim}$120 THz (280 to 2500 nm),
into 1200 bins preserves spectral features and does not hinder convergence.
Each of these $n$ bins represents the integrated photon flux across the
frequency bin. Thus, the photon absorption is given by

\begin{equation}\label{eqn:jl}
    J_{L}=\sum_{i=1}^{n}I_{sol,i}(1-e^{-\tau_{i}})
\end{equation}
where $\tau_{i}$ is the optical depth of the pigment over the frequencies in
bin $i$, and $I_{sol,i}$ (photons m$^{-2}$ s$^{-1}$) is the integrated incident
photon flux per unit area for the given frequency bin. Here, $e^{-\tau_{i}}$ is
the transmittance, so equation \ref{eqn:jl} is the familiar description of the
attenuation of external photons travelling through a medium (the cell). In this
model, the photon flux with respect to frequency distribution is considered
appropriate, since photosynthesis is a quantum process and frequency rather
than wavelength is relevant due to the Kuhn-Reiche-Thomas sum rule for dipole
transitions between electronic states \citep[e.g.][]{bjorn_why_1976}.

Similar to $J_{L}$, the thermal excitation rate, $J_{D}$, is calculated across
the same $n$ frequency bins and given by

\begin{equation}\label{eqn:jd}
    J_{D}=\sum_{i=1}^{n} \tau_{i} \cdot I_{bb,i}
\end{equation}
where $I_{bb,i}$ (photons m$^{-2}$ s$^{-1}$) is the integrated blackbody photon
flux from the pigments at 295 K (room temperature, following
\citet{marosvolgyi_cost_2010}) per unit area for the frequency range of bin
$i$. 

The definition of $J_{D}$ in equation \ref{eqn:jd} arises from two assumptions.
First, it is assumed that the cell is in local thermodynamic equilibrium.
Second, it is assumed that there are a sufficient number of pigments in the
cell that the number of excited pigments does not appreciably deplete the
number of unexcited pigments \citep{ross_thermodynamics_1967}.  Under these
assumptions, all photons emitted by the cell due to thermal excitation will be
reabsorbed within the cell so the rate of thermal pigment excitation is equal
to the rate of thermal emission. The total rate of thermal emission from the
pigments in a cell is given by the number of pigments times their
frequency-dependent emissivity times the frequency-dependent blackbody flux. If
we assume the cell has some pigment density, $n_{a}$ (pigments m$^{-3}$), some
path length $s$ (m), and an absorption cross section $\sigma_{i}$ (m$^{2}$
pigment$^{-1}$), then the thermal emissivity is given by $n_a \cdot s \cdot
\sigma_{i}$ multiplied by the blackbody flux of the pigments, $I_{bb,i}$.
Optical depth is defined by $\tau_{i} = n_a \cdot s \cdot \sigma_{i}$
\citep[e.g.][\S2.4.2.2]{catling_atmospheric_2017}, which
gives equation \ref{eqn:jd}.

In equations \ref{eqn:jl} and \ref{eqn:jd}, we use the optical depth,
$\tau_{i}$, to calculate pigment excitation rates. This differs from the
derivation of \citet{marosvolgyi_cost_2010}, which defined a dimensionless
absorption cross-section, $\sigma_{i}^{*}$, in place of $\tau_{i}$ (the $*$
superscript is added here to denote the dimensionless parameter from
\citet{marosvolgyi_cost_2010}). To derive the model as described in the
supplemental material of \citet{marosvolgyi_cost_2010}, $\sigma_{i}^{*}$ is
defined by the number of dipoles per frequency bin, $g_{i}$.  That is, with
units dropped, $\sigma_{i}^{*}=g_{i} \cdot h \nu_{i} \cdot B/c = d_{i}^{*}
\cdot h \nu_{i}$ for Planck constant $h$, frequency $\nu_{i}$, Einstein
coefficient $B$, and speed of light $c$. Optical depth, which depends on
$\sigma_{i}$, can be similarly defined by $\tau_{i} = g_{i} \cdot h \nu_{i}
\cdot B/c \cdot n_{a}s = d_{i} \cdot h \nu_{i}$. The model solves for the
optimal number of dipoles via $d_{i}^{*}$, or equivalently $d_{i}$, that
maximizes photosynthetic energy production in a cell, as discussed below. Thus,
using $\tau_{i}$ instead of $\sigma_{i}^{*}$ is purely semantic and the model
derivation is unchanged from the method presented in
\citet{marosvolgyi_cost_2010}.

From the power input to the cell, $P_{in}$, which depends on $\tau_{i}$ through
$J_{L}$ and $J_{D}$, the power output from the photosynthetic apparatus is
defined as $P_{out}$. $P_{out}$ represents the total chemical energy stored by
the cell, which accounts for energy losses from $P_{in}$ to heat and reemission
of captured photons, so $P_{out} < P_{in}$. Following
\citet{marosvolgyi_cost_2010}, a simple hyperbolic relationship between
$P_{in}$ and $P_{out}$ is assumed and that $P_{out}$ saturates at some power
output, $P_{sat}$. With these assumptions, $P_{out}$ is given by

\begin{equation}\label{eqn:pout}
    P_{out} = \frac{1}{1/P_{in} + 1/P_{sat}}.
\end{equation}
The saturation term in equation \ref{eqn:pout}, $P_{sat}$, is necessary to
account for the diminishing energy output per pigment as a cell becomes
increasingly absorbent. For example, a black cell that absorbs all incoming
photons will not harvest more photon energy by creating additional pigments and
has thus reached the saturation point for power output.

The proteins required to build these photosynthetic pigments and the subsequent
chemical storage of the captured energy must have some cost to the cell.
Following \citet{marosvolgyi_cost_2010}, we define $C_{P_{in}}$ as the
dimensionless fractional energetic cost to the cell to create the
light-harvesting machinery. Similarly, $C_{P_{out}}$ is defined as the
fractional cost to the cell to create the mechanism necessary to store the
captured photon energy. The remaining fractional energy, that is the relative
energy not spent on light harvesting and subsequent storage, is spent driving
cellular growth, $C_{G}$, or $C_{G} = 1 - C_{P_{in}} - C_{P_{out}}$. 

With the cellular costs $C_{P_{in}}, C_{P_{out}}$, and $C_{G}$ defined, the
growth power available to the cell, $P_{G}$, is simply defined as

\begin{equation}\label{eqn:pg}
    P_{G} = P_{out} \cdot C_{G} = P_{out} \cdot (1 - C_{P_{in}} - C_{P_{out}}).
\end{equation}
We want to find the number of dipoles in each frequency bin, that is the
$d_{i}$, that maximize $P_{G}$. This maximum will occur when the gradient of
$P_{G}$ with respect to $d_{i}$ is 0, i.e. $\partial P_{G} / \partial d_{i}=0$.
The derivation of $\partial P_{G} / \partial d_{i}$ is carried out in detail in
the supplemental material of \citet{marosvolgyi_cost_2010} and produces $n$
equations (recall that $n$ is the number of frequency bins) with the form 

\begin{equation}\label{eqn:full}
    I_{sol,i}\cdot h\nu_{i}\cdot e^{-\tau_{i}} =  \frac{kT\cdot e^{\mu/(kT)}
    \cdot h\nu_{i}}{\mu + kT} \cdot I_{bb,i} + \frac{P_{in}}{(\mu + kT) \cdot
\sum_{j=1}^{n}\tau_{j}/(h\nu_{j})} \cdot \frac{C_{P_{in}}}{C_{P_{in}} + C_{G}}.
\end{equation}
In equation \ref{eqn:full}, $h$ is the Planck constant ($6.626\times10^{-34}$ J
s) and $C_{P_{in}}/(C_{P_{in}} + C_{G})$, which we define as
$C=C_{P_{in}}/(C_{P_{in}} + C_{G})$, is the \textit{relative cost parameter}
that represents the ratio of the cost of the light harvesting machinery to the
sum of the light harvesting and fractional growth energies
\citep{marosvolgyi_cost_2010}. $C$ is a free parameter in the model and $0 \leq
C \leq 1$. When $C=0$, there is no cost to the cell to create photosynthetic
pigments leaving all energy to drive growth. When $C=1$, all absorbed energy is
spent creating the photosynthetic pigments and no energy is available for
growth.

The left-hand side of equation \ref{eqn:full} is the transmitted photon
energy and is balanced by two terms on the right-hand side (RHS). The first
term on the RHS of equation \ref{eqn:full} is the scaled blackbody flux of the
pigment. This term sets the frequency below which pigments would emit more
energy than they absorb and results in an abrupt transition to a transmittance
of 1 below the cutoff frequency. This frequency cutoff is analogous to the
bandgap in photovoltaic semiconductors \citep{marosvolgyi_cost_2010} and
depends on the incident stellar photon flux.

\begin{figure}[ht]
    \centering
    \makebox[\textwidth][c]{\includegraphics[width=0.8\textwidth]{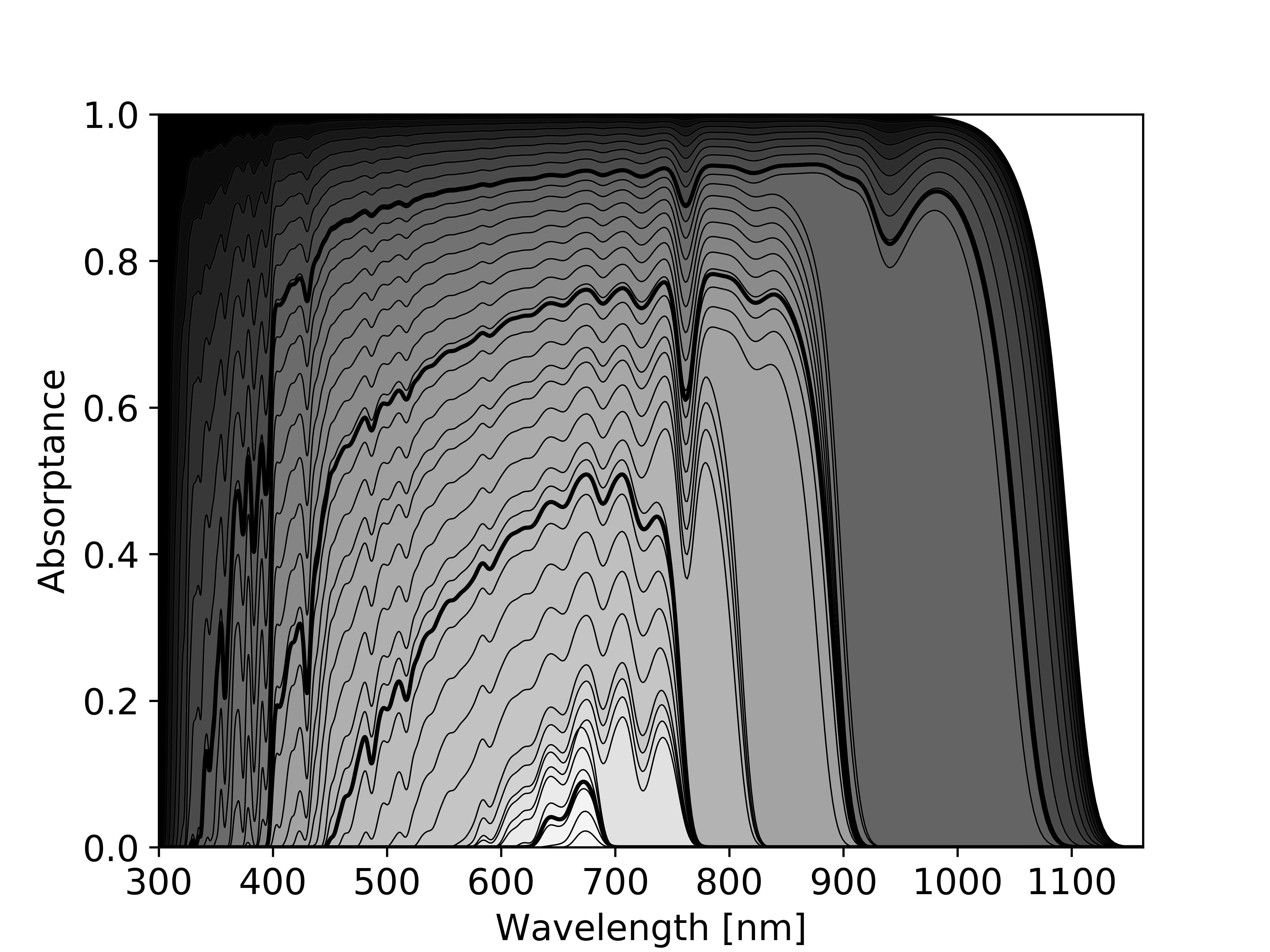}}%
    \caption{The optimal pigment absorptance spectrum for different cost
        parameters around the Sun, a G2V star. The horizontal axis shows
        wavelength in nm, the vertical axis shows pigment absorptance. Each
        black curve shows the predicted pigment absorptance for the given cost
        parameter, $C$ from equation \ref{eqn:full}. The value of $C$ is
        indicated by the shading where darker shading indicates lower cost.
        Black shading corresponds to no cost, i.e.  $C=0$; shading approaches
        white as the cost increases, i.e. $C\rightarrow1$.  The bold black
        contours show, from top to bottom, a $C$ value of 0.25, 0.5, 0.75, and
        0.962, respectively.  The predicted absorptance spectrum was smoothed by
        convolution with a 10 nm wide Gaussian function. As $C$ increases, the
        optimal absorptance spectra focuses to spectral regions where
        high-energy photons are most abundant. This figure recreates the
        results of \citet{marosvolgyi_cost_2010}, see their Figure 2.  }
    \label{fig:varied_cost_g2v}
\end{figure}

The second term on the RHS of equation \ref{eqn:full} imposes an input energy
threshold on the frequencies at which a pigment can absorb
\citep{marosvolgyi_cost_2010}.  This term is spectrally constant but depends on
$C$. As $C \to 0$, this term goes to 0 causing the spectral bins below the
bandgap frequency to absorb all available photons. As $C$ increases from 0,
only the spectral bins where energy input to the cell relative to thermal
emission is largest continue to absorb.  These high-energy input bins represent
the most efficient frequencies for absorption as we assume pigment creation
costs have no spectral dependence. Thus, $C$ can analogously be considered a
measure of pigment efficiency. As $C$ increases, pigment absorption in
inefficient parts of the spectrum is precluded and only remains at frequencies
where stellar photon energy is abundant.  This is readily seen in Figure
\ref{fig:varied_cost_g2v}, which shows the predicted absorptance spectrum of an
optimized cell from equation \ref{eqn:full} for the modern Earth around the Sun
with various $C$ values between 0 and 1.

Equation \ref{eqn:full} is a fixed-point equation that can be solved for
$\tau_{i}$ with iterative mapping. For a given relative cost parameter, $C$, an
initial guess for each $\tau_{i}$ is provided, then equation \ref{eqn:full} is
iterated until convergence is reached. For values of $C$ near 0 or 1, the
initial guess for each $\tau_{i}$ must be close to the solution for the model
to converge.  Therefore, we initially run the model with an intermediate value
for $C$ (such as $C=0.5$) and initialize each $\tau_{i}$ to 1 then iterate
equation \ref{eqn:full} to convergence. With the converged solution for each
$\tau_{i}$ at the intermediate $C$ value, we iterate over $C$ to $C=1$ or $C=0$
using the previously converged solution for each $\tau_{i}$ as inputs to
equation \ref{eqn:full} for the next value of $C$. In this way, equation
\ref{eqn:full} converges for all $C$ values between 0 and 1, including the end
points.

The effect of changing $C$ on the optimal pigment absorbance spectrum from
equation \ref{eqn:full} is shown in Figure \ref{fig:varied_cost_g2v}. As $C$
increases, equation \ref{eqn:full} predicts pigment absorption should focus on
spectral regions where the balance between photon absorption and thermal
emission maximizes energy input to the cell. Thus, pigment absorbance spectra
should jump across atmospheric absorption bands. This is readily seen in Figure
\ref{fig:varied_cost_g2v}, where jumps in predicted pigment absorptance
correspond to atmospheric absorption features, such as those of O$_{2}$ at 688
and 761 nm and from H$_{2}$O at 720, 820, and 940 nm
\citep{hill_absorption_2000, kiang_spectral_2007}. 

As wavelength increases, blackbody emission of the organism becomes
increasingly important. Even in the scenario $C=0$, no pigment
absorption is predicted beyond ${\sim}1100$ nm for the modern Earth around the
Sun. Beyond this wavelength, there are insufficient photons to make up for the
emissions from the pigments. Thus, in this model, pigments cannot absorb beyond
${\sim}1100$ nm without a loss of energy. This limit agrees with extant Earth
organisms and theoretical calculations of the long-wavelength limit of
light-driven energy production \citep{marosvolgyi_cost_2010}. For different
stellar types this limit will change based on the incident flux of infrared
photons, as shown in Appendix \ref{appendix:cost_variations}. We note that
other limitations on the long-wavelength limit for oxygenic photosynthesis may
exist \citep[e.g.][]{kiang_spectral_2007, lehmer_productivity_2018}, but no
such limits are imposed in the model presented here.

The relative cost parameter, $C$, can be freely tuned in the model. However, a
relative cost of $C=0.962$, indicating only the most efficient frequency bins
should absorb photons, reproduces the red absorption feature of spinach
chloroplasts. Furthermore, the same $C$ applied to the ambient spectra in a
muddy water column reproduces the absorption position and shape of the
\textit{Rhodobacter sphaeroides} chromatophores \citep{marosvolgyi_cost_2010}.
Following \citet{marosvolgyi_cost_2010}, we assume $C=0.962$ is applicable to
photosynthetic organisms generally. However, lab measurements of additional
photosynthetic organisms are necessary to know if $C=0.962$ is universally
applicable for Earth-based life or if a range of $C$ values may be appropriate.

\begin{figure}[htb!]
    \centering
    \makebox[\textwidth][c]{\includegraphics[width=0.9\textwidth]{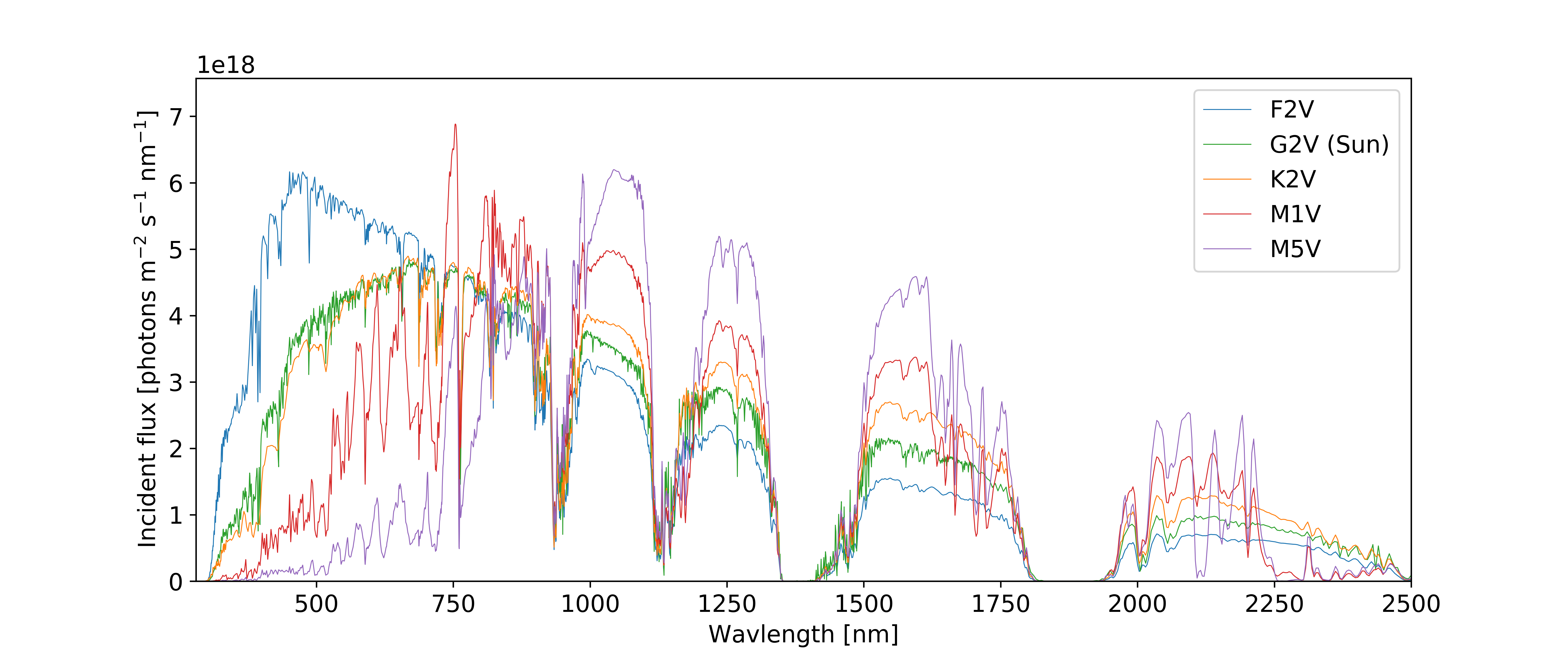}}%
    \caption{The surface spectral photon flux for planets with \changed{modern
        Earth-like CO$_2$, N$_2$, O$_2$, and surface-to-atmosphere reactive gas
        fluxes that are equilibrated with the spectral radiation of different
        parent stellar types (see text for details)}.  The horizontal axis shows
        photon wavelength in nm, the vertical axis shows the incident photon flux
        per unit area, per second, per nm. Each colored contour shows the surface
        photon flux for a different stellar type, as labelled. For each stellar
        type, the total stellar flux \changed{corresponds to a mean planetary
        surface temperature of 288 K}.  The photon flux for the Sun is from
        \citet{astm_reference_2020}. All other spectra are from
\citet{kiang_spectral_2007}.}
    \label{fig:spectra}
\end{figure}

As inputs to the model, we use the standard solar irradiance spectrum for the
modern Earth at sea level from \citet{astm_reference_2020}. \changed{We also
    consider simulated surface spectra from \citet{kiang_spectral_2007} around
    F2V, K2V, M1V, and M5V type stars. For simulated spectra around other
    stars, the atmosphere was assumed to be 1 bar with O$_2$, CO$_2$, and N$_2$
    mixing ratios equivalent to modern Earth's. For fixed O$_2$, the
atmospheric compositions were photochemically evolved to equilibrium based on
the incident stellar irradiance spectrum (see \citet{kiang_spectral_2007} for
details)}. The spectra considered in this work are shown in Figure
\ref{fig:spectra}.

\section{Results}\label{sec:results}

\begin{figure}[htb!]
    \centering
    \makebox[\textwidth][c]{\includegraphics[width=0.8\textwidth]{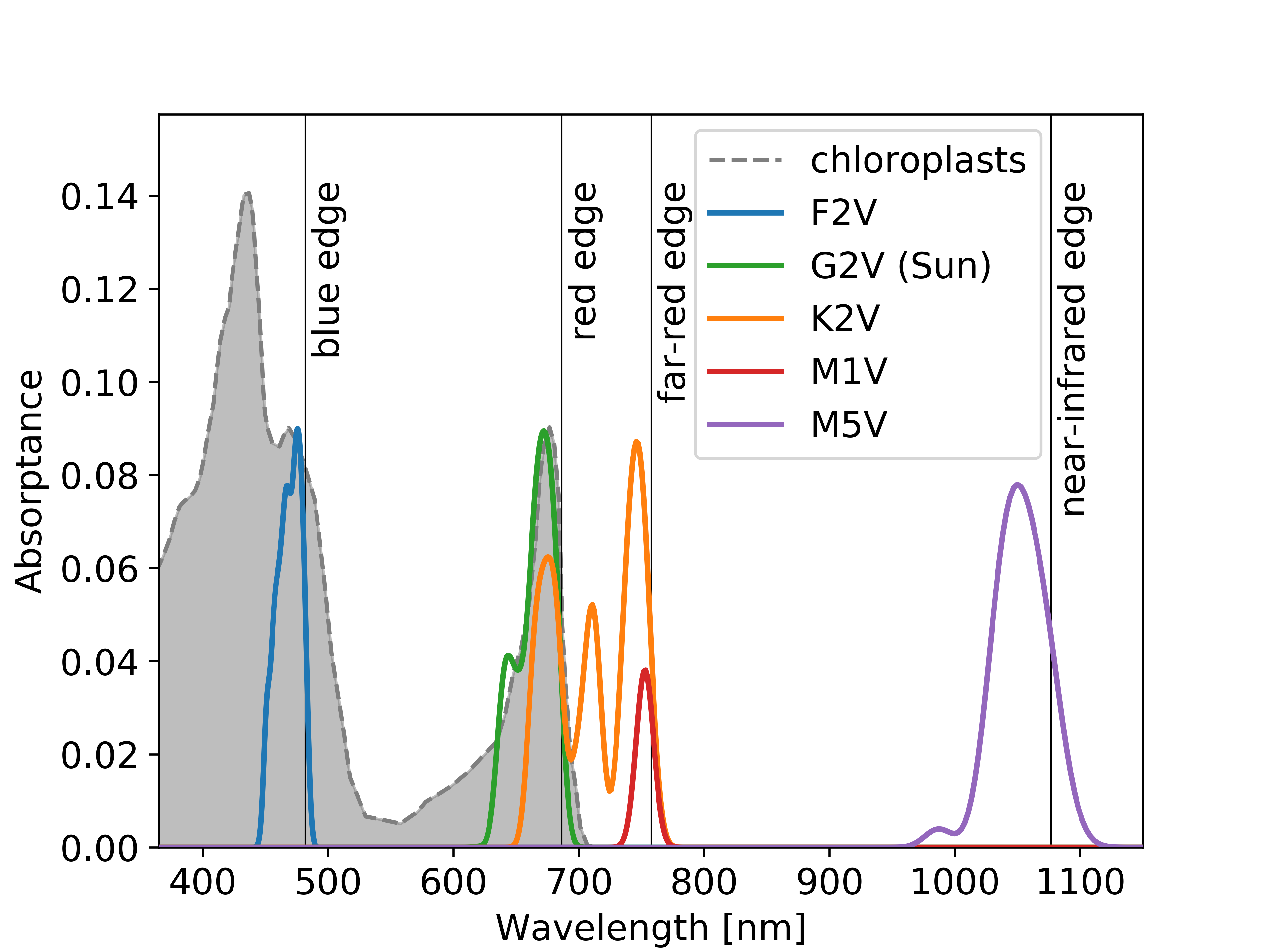}}%
    \caption{The optimal pigment absorbance spectrum for each stellar type from
    our pigment optimization model (see Section \ref{sec:methods}). The
horizontal axis shows wavelength in nm. The vertical axis shows the predicted
absorptance of pigments from the model. Each colored contour represents a
different stellar type, as labelled. The shaded gray region and dashed gray
line show the absorptance of spinach chloroplasts from
\citet{marosvolgyi_cost_2010}, arbitrarily normalized to match the vertical
extent of the optimal pigment profile for Earth. The absorbance spectra were
smoothed by convolution with a 10 nm wide Gaussian function, following
\citet{marosvolgyi_cost_2010}. The model predicts the wavelength of peak
absorption and the corresponding absorption shape around that peak. The total
absorptance of an organism is not reflected by the vertical axis. An organism
could produce additional pigments to boost total absorptance, but the optimized
absorption peak and shape would remain the same. The position of and shape of
\chla{} absorption at $\sim$680 nm (dashed gray contour) is predicted in both
location and shape by the model (green contour). The reflectance edges that
future telescopes might search for are shown by the vertical black lines and
corresponding labels.}
    \label{fig:absorption}
\end{figure}

From the spectra in Figure \ref{fig:spectra}, Figure \ref{fig:absorption} shows
the predicted absorbance spectrum for optimal pigments around each stellar
type. The gray shaded region shows the absorbance spectrum of spinach
chloroplasts from \citet{marosvolgyi_cost_2010}, which matches the predicted
absorbance spectrum in the $Q_y$ band of \textit{in vivo} \chla{} on Earth
(green contour in Figure \ref{fig:absorption}). The green contour in Figure
\ref{fig:absorption} reproduces the results of \citet{marosvolgyi_cost_2010}.
It is important to note in Figure \ref{fig:absorption} that the absolute
absorptance value predicted by the model is not necessarily representative of
the total absorptance of an organism. The model predicts an optimal absorption
wavelength and profile over wavelength, but an organism could produce multiple
copies of the optimal pigment, which would alter the absolute absorptance of
the organism.

\begin{table}
\begin{center}
 \caption{\label{tab:abs_peaks} The model predicted absorptance peaks for each
          stellar type shown in Figure \ref{fig:absorption}.}
 \begin{tabular}{l c} 
     Stellar Type & Absorptance Peaks {[}nm{]}\\ [0.5ex] 
 \hline\hline
    F2V & 468, 476\\
    G2V (Sun) & 644, 672\\
    K2V & 675, 711, 746\\
    M1V & 753\\
    M5V & 987, 1050\\
 \hline
 \end{tabular}
\end{center}
\end{table}

The optimized absorptance peaks shown in Figure \ref{fig:absorption} are listed
in Table \ref{tab:abs_peaks} and agree with the qualitative predictions from
\citet{kiang_spectral_2007}. In addition to predicting the absorptance peaks, we
also predict the shape of the absorbance spectra around each peak and,
depending on stellar type, multiple absorptance peaks. The shapes of the
absorptance peaks depend on the incident stellar flux, atmospheric absorption,
and the relative cost parameter. The broad absorptance peak predicted for an
M5V star arises due to abundant available photons near 1050 nm with limited
atmospheric attenuation around that peak. Stronger atmospheric absorption
features near the absorptance peaks for the other stellar types restrict the
width of the predicted absorptance peaks.

\begin{figure}[h]
    \centering
    \makebox[\textwidth][c]{\includegraphics[width=0.9\textwidth]{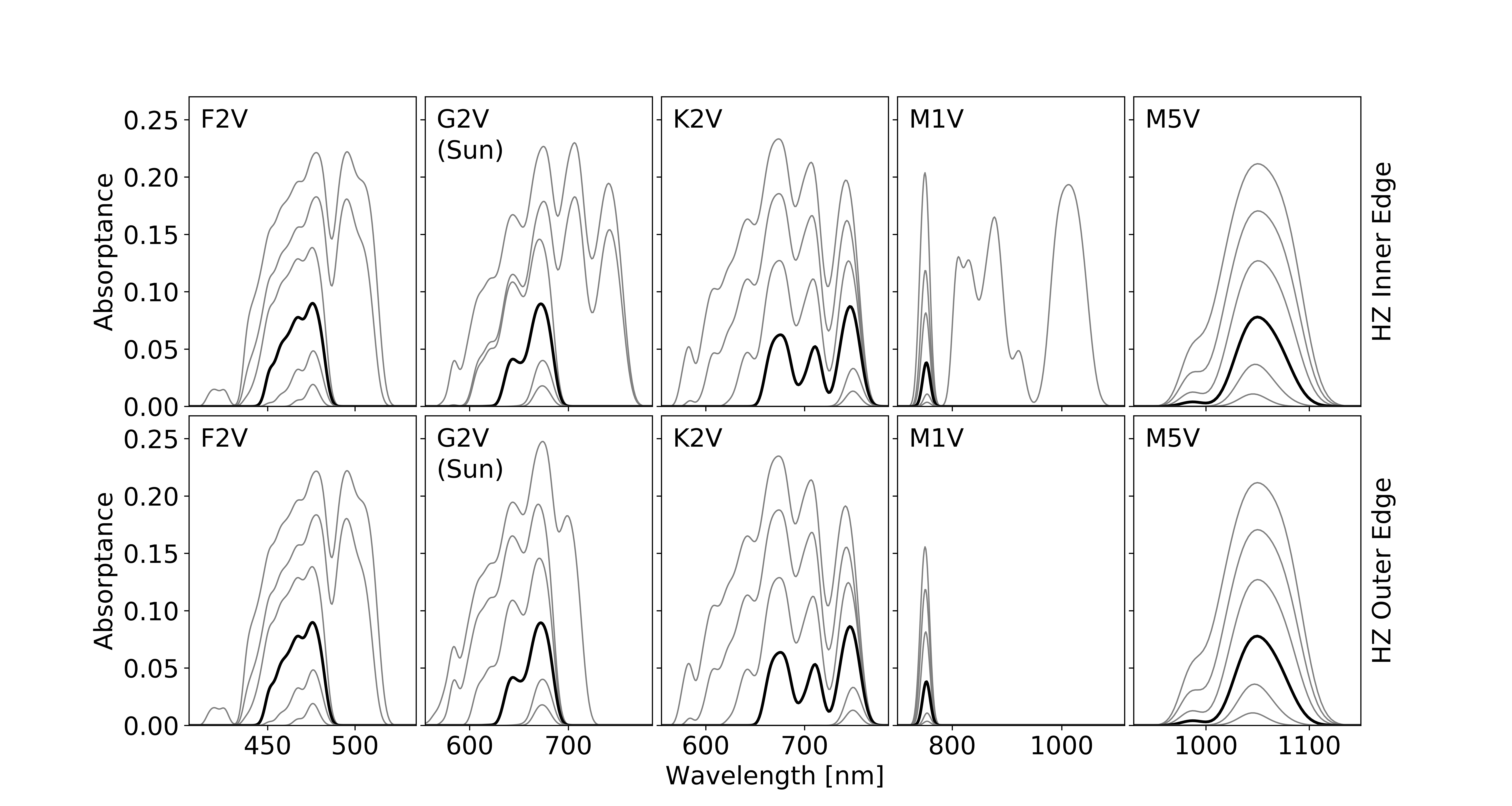}}
    \caption{The optimal pigment absorptance predicted by the model for each
        stellar type \changed{for irradiance intensities approximately spanning
        the} inner and outer edges of the habitable zone (HZ)
        for various relative cost parameters, $C$.  From top to bottom, the
        contours in each figure correspond to a $C$ value of: 0.90, 0.92, 0.94,
        0.962, 0.98, 0.99. The 0.962 contour is shown in bold black while all
        other contours are in gray.  The top row of plots shows the optimal
        pigment absorptance profile at the inner edge of the HZ, the bottom row
        at the outer edge. The flux values used for the inner and outer edge of
        the HZ are given in Table \ref{tab:hz_limits}. For $C\geq0.94$ the
        predicted absorptance profile is the same throughout the HZ for all
        stellar types.  At lower $C$ values, the optimal absorptance profile
        can change between the inner and outer edges of the HZ due to changes
        in total incident flux, as shown by the $C=0.90$ contour for the M1V 
        star.}
    \label{fig:hz_diff}
\end{figure}

The optimal absorbance spectra depend on the relative cost parameter, $C$, and
the total available stellar photon flux for absorption, as demonstrated in
Figure~\ref{fig:hz_diff}. As gray contours, Figure~\ref{fig:hz_diff} shows the
model predicted optimal absorbance spectra for 6 different values of $C \in
\{0.90,0.92,0.94,0.962,0.98,0.99\}$ at the inner and outer edges of the
habitable zone (HZ). The $C=0.962$ contour is highlighted in black. As a
first-order approximation, the photon fluxes available at the inner and outer
edges of the HZ for each stellar type are calculated by linearly scaling the
photon fluxes from Figure~\ref{fig:spectra} by the HZ flux limits in
Table~\ref{tab:hz_limits}. \changed{The HZ flux limits considered in
Figure~\ref{fig:hz_diff} do not take into account environmental and atmospheric
changes that may occur by moving an Earth-like planet between the inner and
outer edges of the HZ. Thus, the results shown in Figure~\ref{fig:hz_diff} are
only illustrative of how changes in total stellar flux alone could influence
optimal pigment absorption profiles.}

In the outer HZ, the decreased total stellar flux can push the optimal
absorbance peak wavelength toward shorter wavelengths when $C$ is low. This is
readily seen for the $C=0.90$ curve of the M1V star in Figure
\ref{fig:hz_diff}. In the inner HZ of this star at $C=0.90$, pigments are
optimal when absorbing beyond 1000 nm.  However, in the outer HZ, absorption at
1000 nm is precluded as there are insufficient long-wavelength stellar photons
to overcome thermal emission of the pigments. As $C$ increases, optimal pigment
absorption focuses around the most energetically favorable wavelengths and the
change in total stellar flux between the inner and outer limits of the HZ has
limited influence on optimal pigment absorption wavelength, as seen in
Figure~\ref{fig:hz_diff}.

\begin{table}
\begin{center}
 \caption{\label{tab:hz_limits} The normalized incident fluxes, $S$, at the inner and
     outer edges of the habitable zone (HZ) for each stellar type.  The HZ flux
     limits are normalized to the incident flux of the modern Earth
     ($S_{\oplus}=1360$ W m$^{-2}$) and taken from
     \citet{kopparapu_habitable_2013}.}
 \begin{tabular}{l c c c} 
     Stellar Type & Temperature {[}K{]}& Inner Edge $[S/S_{\oplus}]$ & Outer Edge $[S/S_{\oplus}]$\\ [0.5ex]
 \hline\hline
     F2V & 7120 & 1.11 & 0.41\\
     G2V (Sun) & 5780 & 1.01 & 0.34\\
     K2V & 4620 & 0.93 & 0.28\\
     M1V & 3330 & 0.86 & 0.23\\
     M5V & 2670 & 0.84 & 0.21\\
 \hline
 \end{tabular}
\end{center}
\end{table}

The predicted optimal absorbance spectra for each stellar type shows the
influence of atmospheric absorption on pigment optimization. Comparing the
input spectra in Figure \ref{fig:spectra} to the predicted absorbance
wavelengths in Figures \ref{fig:absorption}, we see that atmospheric absorption
features, particularly from H$_{2}$O at 720, 820, 940, and 1130 nm
\citep{hill_absorption_2000, kiang_spectral_2007}, exert a strong pressure on
pigments to absorb in atmospheric windows. In addition, short-wavelength,
high-energy photons in the visible (and just beyond) provide maximum growth
energy for organisms around F, G, K, and even early M stars.  Only around the
coolest M5V star are organisms optimized when absorbing well into the NIR.
These predictions may better inform the search for red edge analogs around
other stars, as discussed below.

See Appendix \ref{appendix:individual_spectra} for plots comparing the
different stellar spectrum and corresponding predictions for optimal pigment
absorptance.

\section{Discussion}\label{sec:discussion}

If we are to detect a sharp change in reflectance spectra from alien,
photosynthetic organisms, i.e., a red edge analog, knowing the exact
wavelengths to observe is critical as the spectral signal may be small
\citep{sagan_search_1993, seager_vegetations_2005, kiang_spectral_2007-1,
omalley-james_vegetation_2018, wang_diurnal_2021}. In this work, the goal of
predicting pigment absorbance spectra is to inform the wavelength ranges future
observations should preferentially consider for detection of a red edge
analog.

On the modern Earth, the red edge in the NIR corresponds to the steep
drop in absorption beyond the optimal peak of \chla{} (gray shaded region in
Figure \ref{fig:absorption}). If photosynthetic organisms around other stars
exhibit similar edge-like spectral reflectance features, we might expect red
edge analogs to occur on the long-wavelength side of the optimal pigment
absorption, similar to the prediction of \citet{tinetti_detectability_2006}.

\changed{The wavelength of optimal pigment absorption will depend on both the
atmospheric composition and the photon flux of the host star. Both stars
and planetary atmospheres evolve with time
\citep[e.g.][]{bahcall_solar_2001, lyons_rise_2014}, so the optimal
wavelength for pigment absorption may also evolve with time. For the Sun,
changes in luminosity with time have not led to significant spectral change
across the photosynthetically active radiation
wavelengths~\citep{claire_evolution_2012}. In addition, from
Figure~\ref{fig:hz_diff}, we see that changes in total stellar flux do not
alter the optimal pigment absorption wavelengths except at low values of $C$;
lower values of $C$ could also result as adaptations to low light, as seen in
the far-red and low-light adapted cyanobacterium \textit{Acaryochloris
marina}~\citep{mielke_efficiency_2011}. Thus, for the Earth through time,
changes in atmospheric composition could regulate optimal pigment absorption
wavelengths.}

\changed{Atmospheric compositions on habitable exoplanets could be very
    different from modern Earth's~\citep[e.g.][]{segura_ozone_2003,
    segura_biosignatures_2005, schwieterman_exoplanet_2018}, which would filter
    radiation and alter the surface spectral
    irradiance~\citep{kiang_spectral_2007} to which optimal wavelengths for
    pigments absorption would be adapted. A relationship between available
    photon fluxes and pigment absorption profiles is well established on the
    modern Earth (see~\nameref{sec:introduction}) and is even found in
    photoautotrophs where accessory pigments rather than the trap pigment
    dominate the spectral light absorbance. For example, the anoxygenic
    photosynthetic bacterium \textit{Blastochloris viridis} has a
    bacteriochlorophyll~\textit{b} trap wavelength at 960 nm where the
    underwater photon flux in this organism's environment is low, while core
    antenna pigments provide higher absorbance at ${\sim}$890 and 1015 nm (the
    latter actually being up the energy hill from the trap) where photons are
more abundant~\citep{kiang_spectral_2007}. The model used here predicts an
optimal pigment absorption profile based on the available photon flux and does
not differentiate between trap and accessory pigments.}

\changed{As mentioned earlier, land plants have abundant accessory pigments
throughout the visible spectrum. Indeed, green plants absorb almost all
visible photons, including ${\sim}$95\% of green photons
\citep[e.g.][]{bjorn_viewpoint:_2009}, with the help of these accessory
pigments.} It is not surprising that green land plants contain these accessory pigments.
If \chla{} bearing organisms were limited to the narrow range of photons around
the absorption peak at ${\sim}$680 nm, say photons between 670 and 690 nm,
${\sim}9\times10^{19}$ photons m$^{-2}$ s$^{-1}$ would be available from the
spectra in Figure \ref{fig:spectra} for the Earth around the Sun. To first
order, this is only $\sim \frac{1}{4}$ of the current photon use by the
terrestrial biosphere ($3.26\times10^{20}$ photons m$^{-2}$ s$^{-1}$)
\citep{field_primary_1998, lehmer_productivity_2018}. Without accessory
pigments, terrestrial, \chla{}-bearing organisms would not be fully exploiting
available light and potential growth. \changed{Instead, land plants appear to have
adapted by adding (or losing) accessory pigments until photon absorption is no
longer growth limiting and nutrient availability, often bioavailable
phosphorous or nitrogen~\citep{tyrrell_relative_1999, reinhard_evolution_2016},
or water availability on land~\citep{porporato_ecohydrology_2002,
porporato_hydrologic_2003}, limits growth.}

In addition to absorption of short-wavelength photons from accessory pigments,
large pigments, such as \chla{}, should have auxiliary, short-wavelength
absorption features \citep[e.g.][]{papageorgiou_fluorescence_2004}. Pigments
like \chla{} must be sufficiently large that their $\pi$-electrons can be
excited by long-wavelength photons at the edge of the visible spectrum
\citep{mauzerall_chlorophyll_1976}. This large structure results in
additional absorption features from higher energy electron orbitals
\citep[e.g.][]{bjorn_viewpoint:_2009}. So the absorption of blue photons by
\chla{}, which provides additional photon harvesting for a \chla{}-bearing
organism similar to an accessory pigment, may be a side-effect of the structure
of \chla{} being tuned to optimally absorb in the red rather than a feature
that was selected for \citep{marosvolgyi_cost_2010}, but see
\citet{arp_quieting_2020} for an alternative explanation.

If alien photoautotrophs have abundant accessory pigments and auxiliary,
short-wavelength absorption features in their pigments, as occur on Earth,
Earth-like planets around the stars considered here would have sufficient
spectral energy to sustain the Earth's extant biosphere. If we assume photons
longer than 300 nm may be used by photosynthetic organisms
\citep{mccree_action_1971}, more than double the current photon use of the
terrestrial biosphere is available for each stellar type shown in Figure
\ref{fig:spectra}. This assumes absorption only occurs between 300 nm and the
optimal wavelength shown in Figure \ref{fig:absorption}. Even if we account for
the reduced quantum yield of the M5V pigment, which may be ${\sim}50\%$ as
productive due to the lower quantum yield of low-energy, long-wavelength
photons \citep{wolstencroft_photosynthesis:_2002, kiang_spectral_2007,
lehmer_productivity_2018}, no spectral energy limitation on growth is
encountered, although the total stellar flux may be a constraint around the
coolest of stars \citep{lehmer_productivity_2018}. Thus, it may be unlikely for
an organism to produce pigments absorbing beyond the optimal pigment absorption
wavelength. Such pigments may need to be large to absorb the longer wavelength
photons \citep{mauzerall_chlorophyll_1976} and thus could be costly for the
cell to build. In addition, these pigments would absorb lower-energy photons
and potentially have a lower quantum yield compared to accessory pigments at
short wavelengths, where photons are still abundant.

The full reflectance spectra of an organism is derived not only from its
pigments. As noted by \citet{kiang_spectral_2007}, cellular structure, and in
the case of land plants, canopy structure and leaf morphology could play
important roles in determining the reflectance spectra and thus the red
edge. These aspects are not considered in our model so the discussion of the
alien red edge analog location should be considered approximate. However, given
the lack of understanding on how these other properties may alter the red
edge around different stars \citep{kiang_spectral_2007}, we propose the
optimal absorption peak is a reasonable initial location to search for a
vegetative red edge analog, as is the case on the modern Earth.

\section{Conclusion}\label{sec:conclusion}
Future exoplanet observations may search for a vegetative red edge
equivalent, a spectral signature due to a sharp slope in the reflectance of
photosynthetic organisms \citep[e.g.][]{sagan_search_1993,
seager_vegetations_2005, kiang_spectral_2007, omalley-james_vegetation_2018,
wang_diurnal_2021}.  The wavelength where a red edge analog occurs on other
planets will likely depend on the stellar type, \changed{as seen in
Figure~\ref{fig:absorption}. Here, we only consider atmospheric
compositions with CO$_2$, N$_2$, and O$_2$ similar to the modern Earth that
are photochemically equilibrated with the incident stellar photon flux (see
\nameref{sec:methods}), but different atmospheric compositions could also
change the wavelength of red edge analogs by altering the availability of
surface photons~\citep[e.g.][]{kiang_spectral_2007, arney_pale_2016}.}
Direct imaging mission concepts, such as NASA's HabEx and LUVOIR, could
constrain \changed{the atmospheric composition and thus} the surface photon
flux of habitable worlds and search for a vegetative red edge analog based
on that measurement, as modeled here.

The exact nature of oxygenic photosynthesis, why certain absorption features
exist and what drives accessory pigment production remains under investigation
\citep[e.g. see reviews][]{bjorn_viewpoint:_2009, bjorn_spectral_2015}.
However, the model described in this work and similar models can explain the
absorption features, at least in part, of numerous extant Earth organisms
\citep{bjorn_why_1976, stomp_colorful_2007, kiang_spectral_2007-1,
milo_what_2009, marosvolgyi_cost_2010}. An edge-like photosynthetic pigment
signature on an extrasolar planet will depend on the context of the parent
star, atmosphere, and evolutionary stage of the planet.  If evolution of
photosynthesis tends always toward maximizing power gain, then the most likely
candidate for a photosynthetic red edge analog on habitable exoplanets occurs at
the optimal absorption peak, as depicted in Figure~\ref{fig:absorption} for a
selection of stellar types.

\section*{Acknowledgements}
We would like to thank NASA's Virtual Planetary Laboratory (grant
80NSSC18K0829) at the University of Washington and the NASA Pathways Program at
the NASA Ames Research Center for funding this work.

\appendix

\section{Supplemental figures}
\subsection{Cost Parameter Variations} \label{appendix:cost_variations}

Figure~\ref{fig:varied_cost_g2v} shows how changes in the relative cost
parameter, $C$, change the optimal absorbance profile predicted from
equation~\ref{eqn:full} for photosynthetic organisms on the modern Earth around
the Sun.  In this section, we reproduce Figure~\ref{fig:varied_cost_g2v} for
the spectra from each stellar type shown in Figure~\ref{fig:spectra}. Around
each stellar type, the optimal absorbance profile jumps atmospheric absorption
features, similar to Figure~\ref{fig:varied_cost_g2v}.  The long-wavelength
limit for absorption, where thermal emission precludes pigment absorption,
changes based on stellar type.

\begin{figure}[ht]
    \centering
    \makebox[\textwidth][c]{\includegraphics[width=0.9\textwidth]{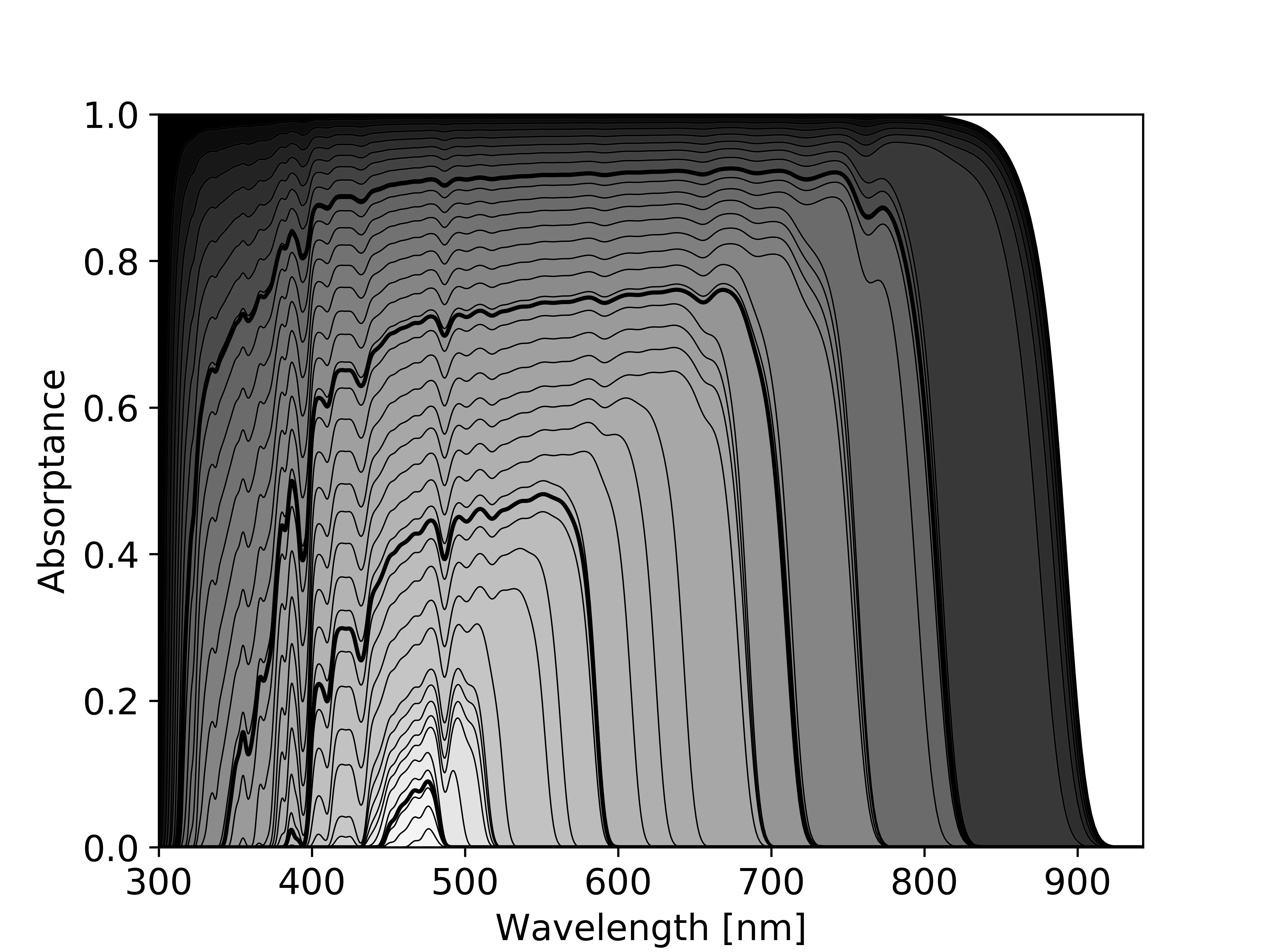}}%
    \caption{
        The optimal pigment absorbance spectrum for different cost parameters
        around an F2V star. See Figure \ref{fig:varied_cost_g2v} (an equivalent
        plot for the Sun) for a full description. }
    \label{fig:varied_cost_f2v}
\end{figure}

\begin{figure}[ht]
    \centering
    \makebox[\textwidth][c]{\includegraphics[width=0.9\textwidth]{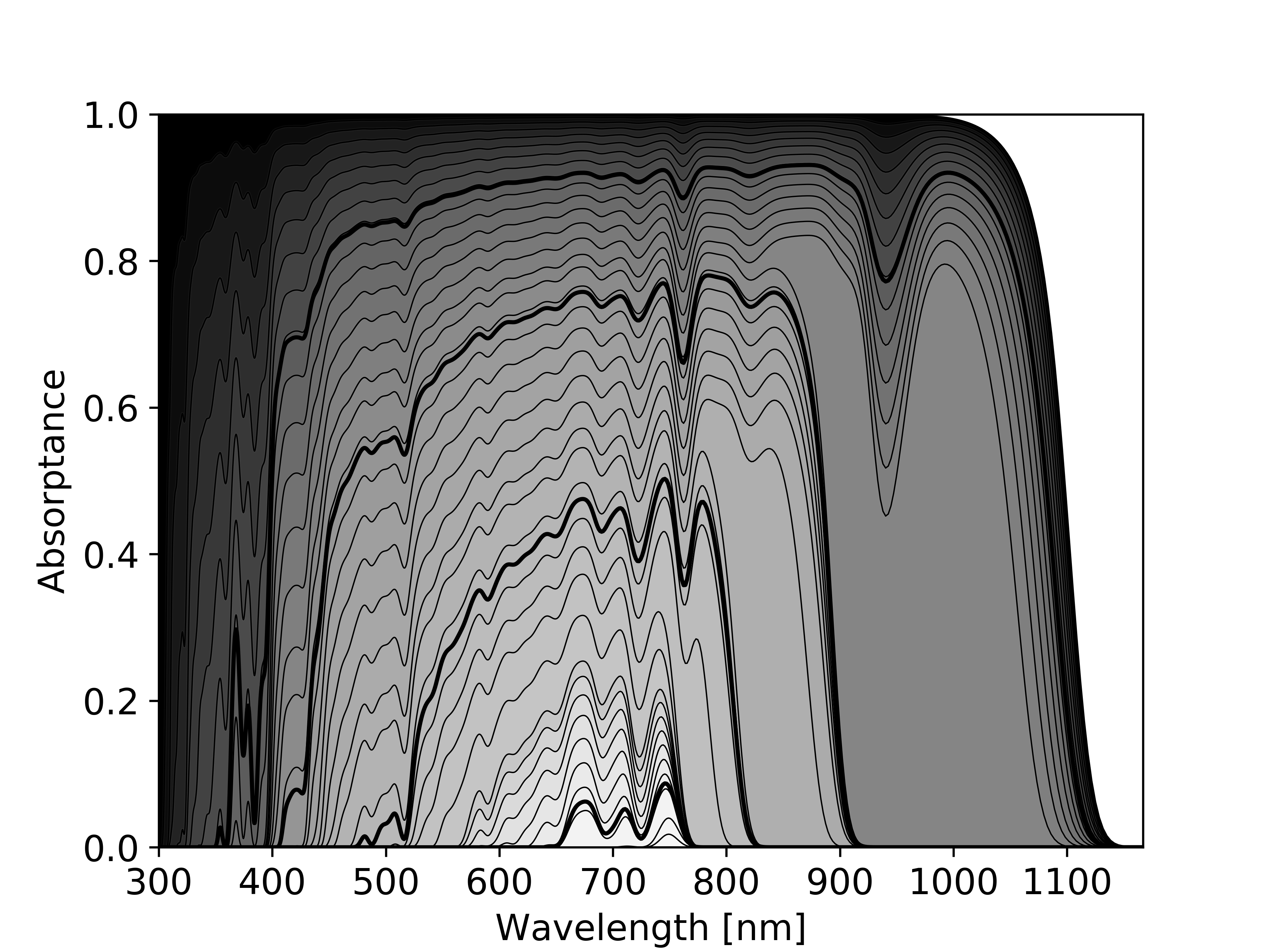}}%
    \caption{
        The optimal pigment absorbance spectrum for different cost parameters
        around an K2V star. See Figure \ref{fig:varied_cost_g2v} (an equivalent
        plot for the Sun) for a full description. }
    \label{fig:varied_cost_k2v}
\end{figure}

\begin{figure}[ht]
    \centering
    \makebox[\textwidth][c]{\includegraphics[width=0.9\textwidth]{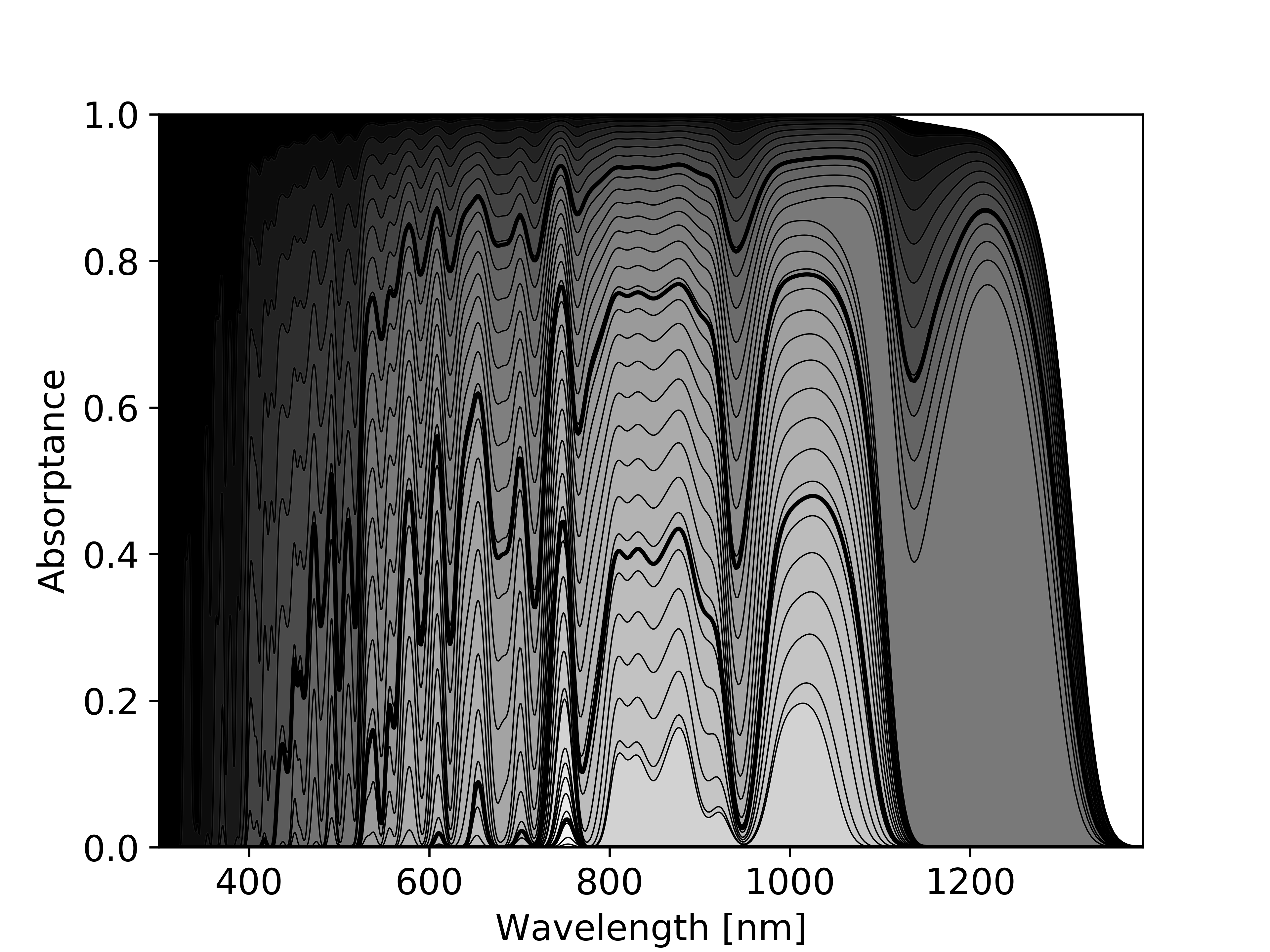}}%
    \caption{
        The optimal pigment absorbance spectrum for different cost parameters
        around an M1V star. See Figure \ref{fig:varied_cost_g2v} (an equivalent
        plot for the Sun) for a full description. }
    \label{fig:varied_cost_m1v}
\end{figure}

\begin{figure}[ht]
    \centering
    \makebox[\textwidth][c]{\includegraphics[width=0.9\textwidth]{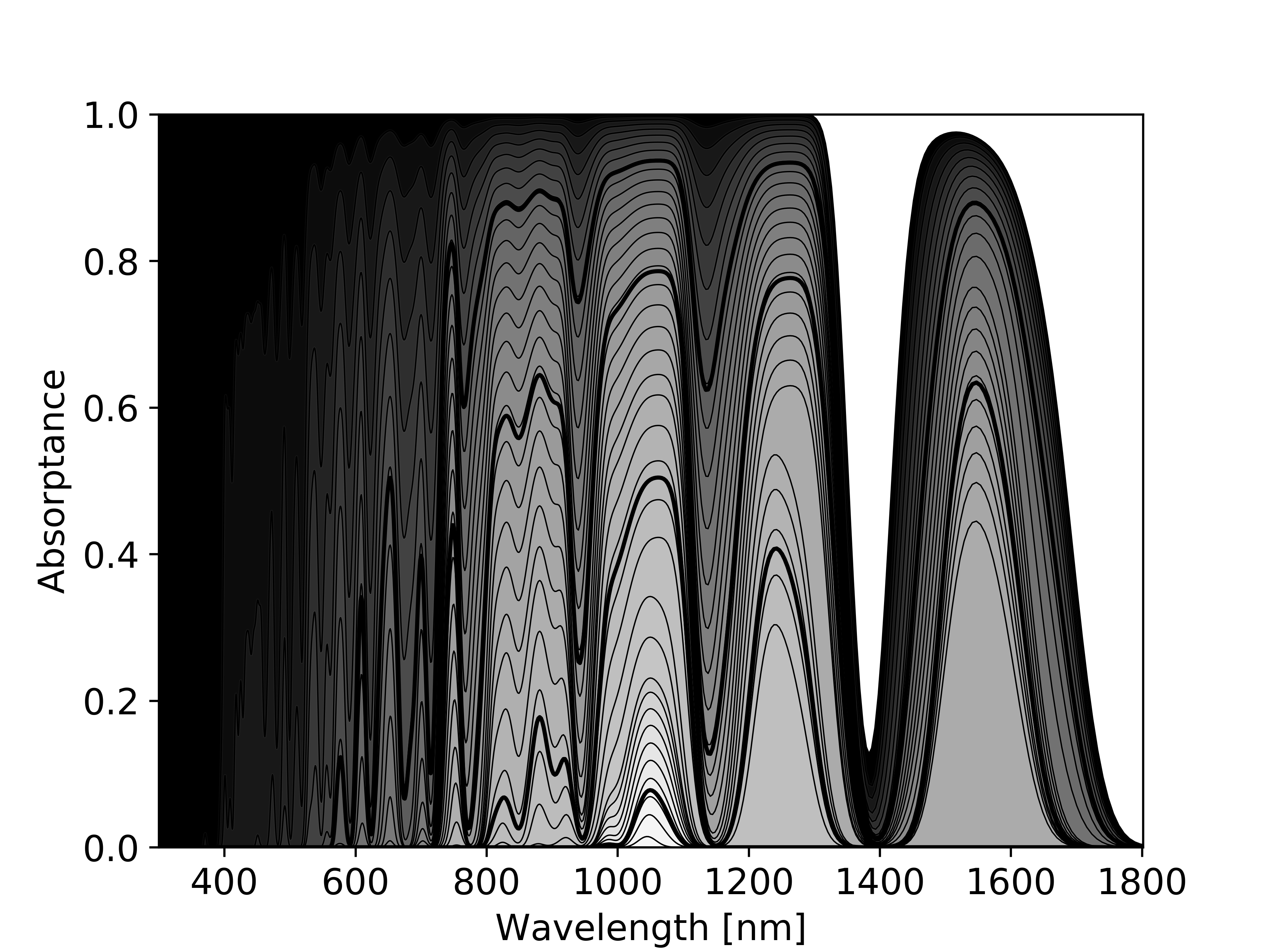}}%
    \caption{
        The optimal pigment absorbance spectrum for different cost parameters
        around an M5V star. See Figure \ref{fig:varied_cost_g2v} (an equivalent
        plot for the Sun) for a full description. }
    \label{fig:varied_cost_m5v}
\end{figure}

\clearpage
\subsection{Individual Spectra}
\label{appendix:individual_spectra}

\begin{figure}[ht]
    \centering
    \makebox[\textwidth][c]{\includegraphics[width=1.0\textwidth]{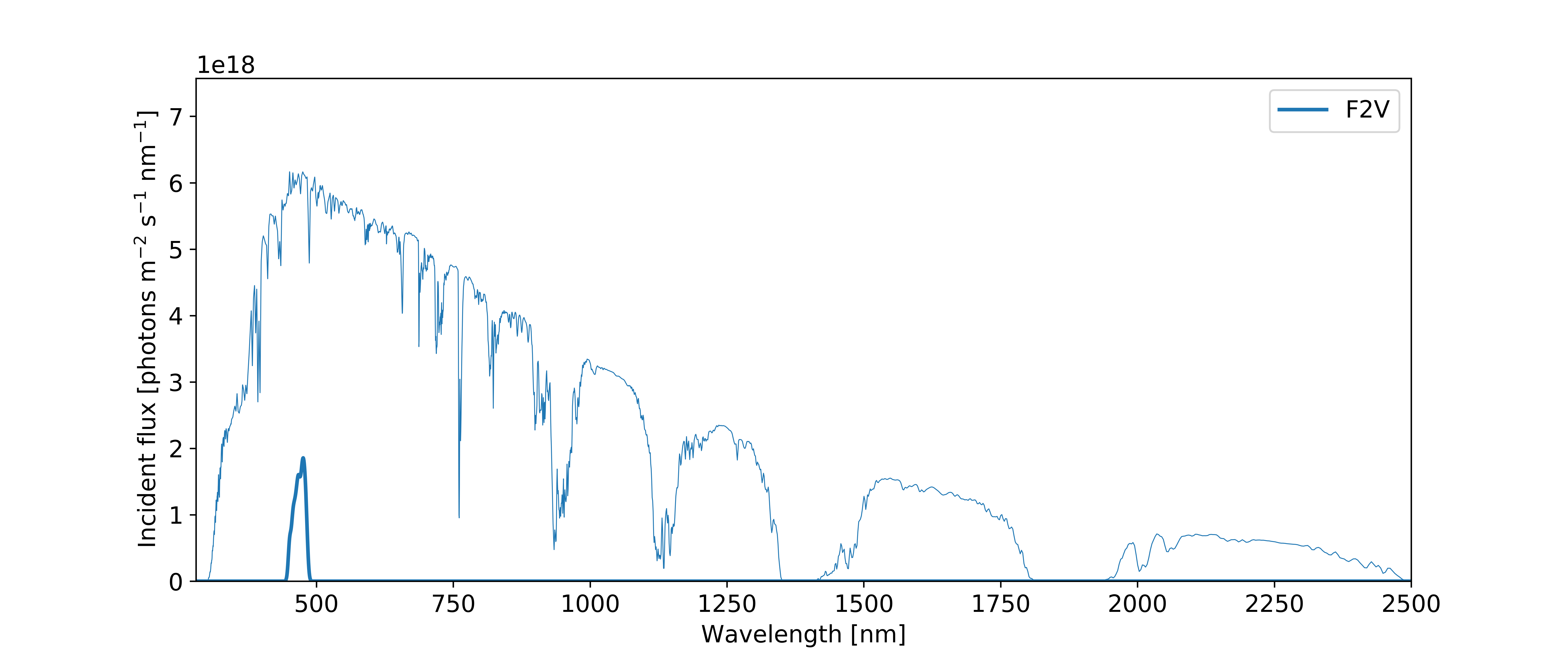}}%
    \caption{ The predicted optimal pigment absorbance spectrum for an F2V
              star. The curves shown here are from Figures \ref{fig:spectra} and
              \ref{fig:absorption}. }
    \label{fig:spectra_and_absorption_f2v}
\end{figure}

\begin{figure}[ht]
    \centering
    \makebox[\textwidth][c]{\includegraphics[width=1.0\textwidth]{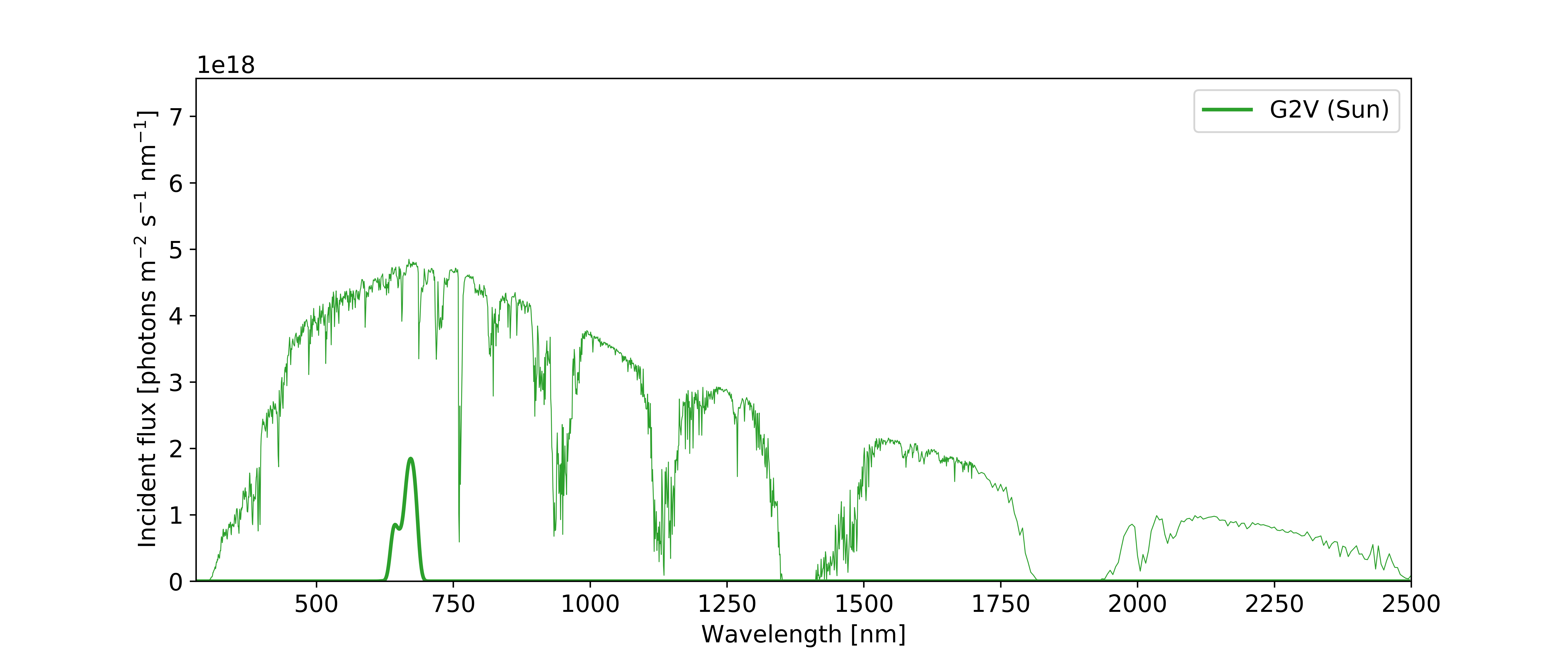}}%
    \caption{ The predicted optimal pigment absorbance spectrum for an G2V
              star. The curves shown here are from Figures \ref{fig:spectra} and
              \ref{fig:absorption}. }
    \label{fig:spectra_and_absorption_g2v}
\end{figure}

\begin{figure}[ht]
    \centering
    \makebox[\textwidth][c]{\includegraphics[width=1.0\textwidth]{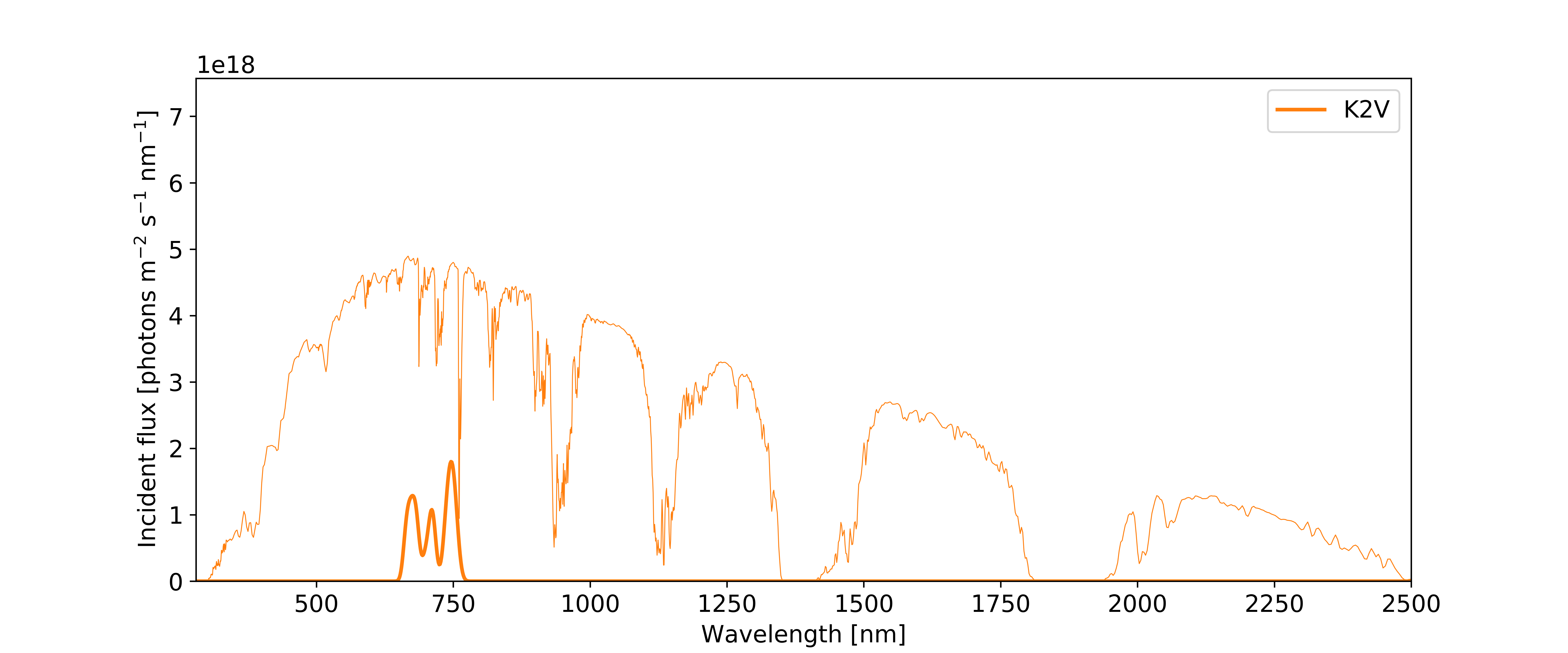}}%
    \caption{ The predicted optimal pigment absorbance spectrum for an K2V
              star. The curves shown here are from Figures \ref{fig:spectra} and
              \ref{fig:absorption}. }
    \label{fig:spectra_and_absorption_k2v}
\end{figure}

\begin{figure}[ht]
    \centering
    \makebox[\textwidth][c]{\includegraphics[width=1.0\textwidth]{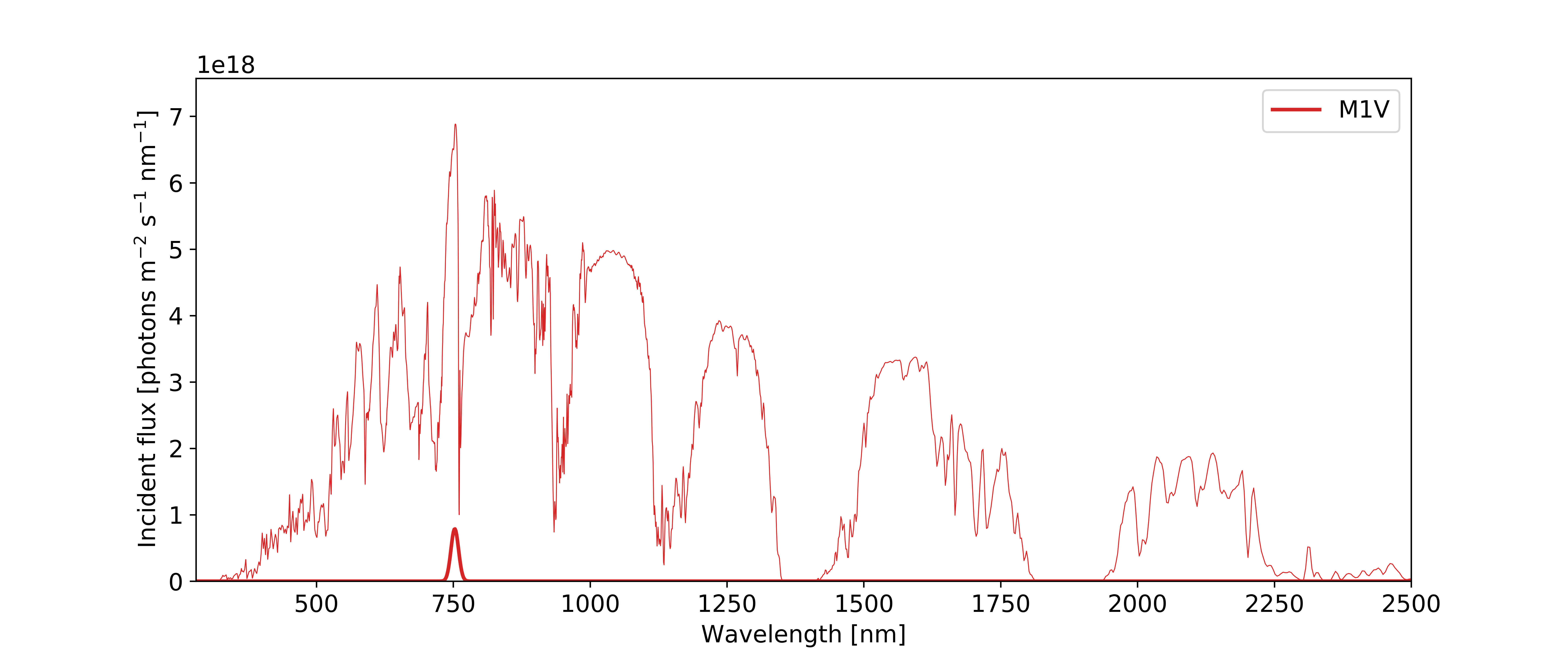}}%
    \caption{ The predicted optimal pigment absorbance spectrum for an M1V
              star. The curves shown here are from Figures \ref{fig:spectra} and
              \ref{fig:absorption}. }
    \label{fig:spectra_and_absorption_m1v}
\end{figure}

\begin{figure}[ht]
    \centering
    \makebox[\textwidth][c]{\includegraphics[width=1.0\textwidth]{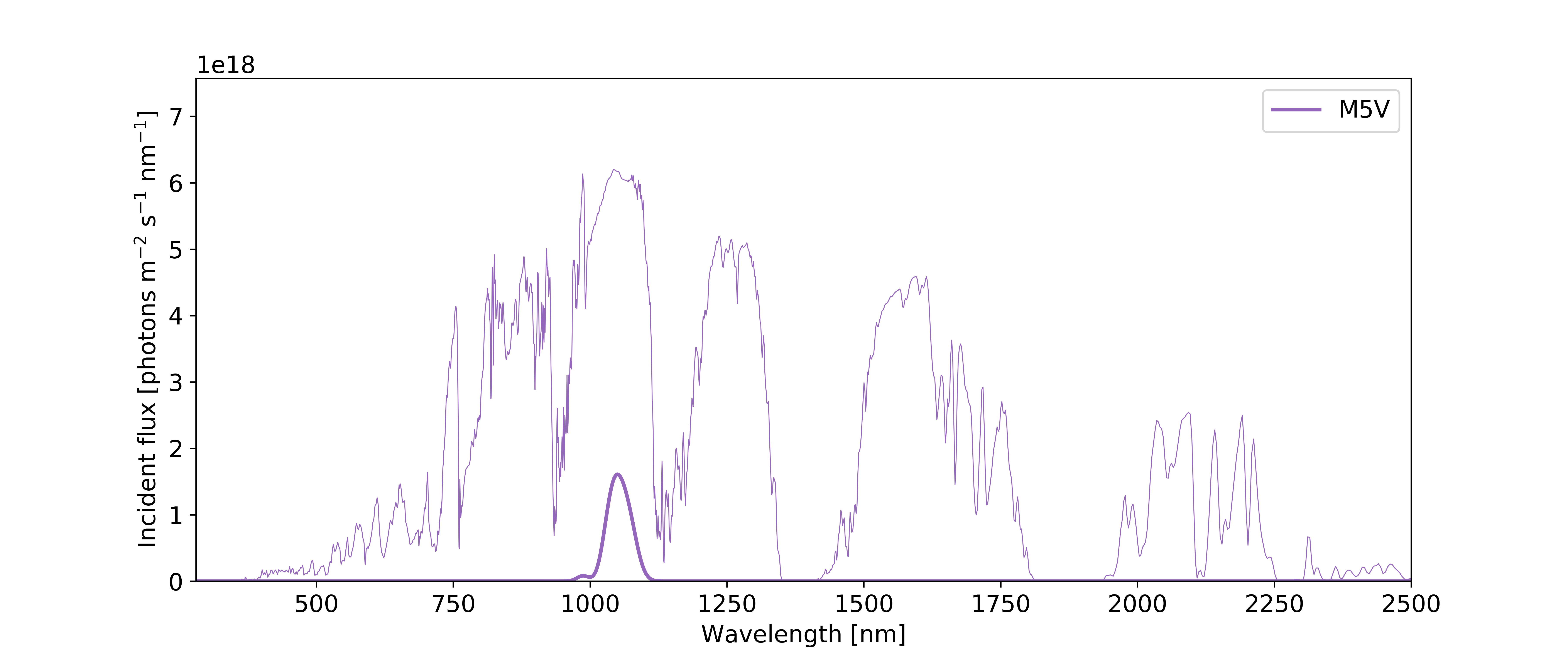}}%
    \caption{ The predicted optimal pigment absorbance spectrum for an M5V
              star. The curves shown here are from Figures \ref{fig:spectra} and
              \ref{fig:absorption}. }
    \label{fig:spectra_and_absorption_m5v}
\end{figure}

The incident photon fluxes for each stellar type considered in this work
(Figure \ref{fig:spectra}) are shown in Figures
\ref{fig:spectra_and_absorption_f2v}, \ref{fig:spectra_and_absorption_g2v},
\ref{fig:spectra_and_absorption_k2v}, \ref{fig:spectra_and_absorption_m1v}, and
\ref{fig:spectra_and_absorption_m5v} with the corresponding prediction for
pigment absorptance from Figure \ref{fig:absorption}. The contours in these
figures are the same contours as shown in Figure \ref{fig:spectra} and Figure
\ref{fig:absorption}, but shown individually for clarity. In each figure, the
vertical scaling of the predicted pigment absorptance profile is arbitrary.
Only the wavelength of the absorptance peak and the absorbance spectra around
that peak are important.

\clearpage

\begin{thebibliography}{}
\expandafter\ifx\csname natexlab\endcsname\relax\def\natexlab#1{#1}\fi
\providecommand{\url}[1]{\href{#1}{#1}}
\providecommand{\dodoi}[1]{doi:~\href{http://doi.org/#1}{\nolinkurl{#1}}}
\providecommand{\doeprint}[1]{\href{http://ascl.net/#1}{\nolinkurl{http://ascl.net/#1}}}
\providecommand{\doarXiv}[1]{\href{https://arxiv.org/abs/#1}{\nolinkurl{https://arxiv.org/abs/#1}}}

\bibitem[{Arney {et~al.}(2016)Arney, Domagal-Goldman, Meadows, Wolf,
  Schwieterman, Charnay, Claire, H{\'e}brard, \& Trainer}]{arney_pale_2016}
Arney, G., Domagal-Goldman, S.~D., Meadows, V.~S., {et~al.} 2016, The {Pale}
  {Orange} {Dot}: {The} {Spectrum} and {Habitability} of {Hazy} {Archean}
  {Earth}, Astrobiology, 16, 873, \dodoi{10.1089/ast.2015.1422}

\bibitem[{Arp {et~al.}(2020)Arp, Kistner-Morris, Aji, Cogdell, Grondelle, \&
  Gabor}]{arp_quieting_2020}
Arp, T.~B., Kistner-Morris, J., Aji, V., {et~al.} 2020, Quieting a noisy
  antenna reproduces photosynthetic light-harvesting spectra, Science, 368,
  1490, \dodoi{10.1126/science.aba6630}

\bibitem[{ASTM(2020)}]{astm_reference_2020}
ASTM. 2020, Reference {Air} {Mass} 1.5 {Spectra} {\textbar} {Grid}
  {Modernization} {\textbar} {NREL}.
\newblock \url{https://www.nrel.gov/grid/solar-resource/spectra-am1.5.html}

\bibitem[{Bahcall {et~al.}(2001)Bahcall, Pinsonneault, \&
  Basu}]{bahcall_solar_2001}
Bahcall, J.~N., Pinsonneault, M.~H., \& Basu, S. 2001, Solar {Models}:
  {Current} {Epoch} and {Time} {Dependences}, {Neutrinos}, and
  {Helioseismological} {Properties}, The Astrophysical Journal, 555, 990,
  \dodoi{10.1086/321493}

\bibitem[{Bj{\"o}rn(1976)}]{bjorn_why_1976}
Bj{\"o}rn, L.~O. 1976, Why are plants green - relationships between pigment
  absorption and photosynthetic efficiency, Photosynthetica, 10, 121.
\newblock \url{http://lup.lub.lu.se/record/134648}

\bibitem[{Bj{\"o}rn \& Ghiradella(2015)}]{bjorn_spectral_2015}
Bj{\"o}rn, L.~O., \& Ghiradella, H. 2015, in Photobiology, ed. L.~O. Bj{\"o}rn
  (New York, NY: Springer New York), 97--117,
  \dodoi{10.1007/978-1-4939-1468-5_9}

\bibitem[{Bj{\"o}rn {et~al.}(2009)Bj{\"o}rn, Papageorgiou, Blankenship, \&
  {Govindjee}}]{bjorn_viewpoint:_2009}
Bj{\"o}rn, L.~O., Papageorgiou, G.~C., Blankenship, R.~E., \& {Govindjee}.
  2009, A viewpoint: {Why} chlorophyll a?, Photosynthesis Research, 99, 85,
  \dodoi{10.1007/s11120-008-9395-x}

\bibitem[{Blankenship(2008)}]{blankenship_basic_2008}
Blankenship, R.~E. 2008, in Molecular {Mechanisms} of {Photosynthesis}
  (Blackwell Science Ltd), 1--10, \dodoi{10.1002/9780470758472.ch1}

\bibitem[{Blankenship(2014)}]{blankenship_molecular_2014}
Blankenship, R.~E. 2014, Molecular {Mechanisms} of {Photosynthesis} (United
  Kingdom: John Wiley \& Sons)

\bibitem[{Catling \& Kasting(2017)}]{catling_atmospheric_2017}
Catling, D.~C., \& Kasting, J.~F. 2017, Atmospheric {Evolution} on {Inhabited}
  and {Lifeless} {Worlds} (New York: Cambridge University Press)

\bibitem[{Claire {et~al.}(2012)Claire, Sheets, Cohen, Ribas, Meadows, \&
  Catling}]{claire_evolution_2012}
Claire, M.~W., Sheets, J., Cohen, M., {et~al.} 2012, The {Evolution} of {Solar}
  {Flux} {From} 0.1 nm to 160 microns: {Quantitative} {Estimates} for
  {Planetary} {Studies}, The Astrophysical Journal, 757, 95,
  \dodoi{10.1088/0004-637X/757/1/95}

\bibitem[{Field {et~al.}(1998)Field, Behrenfeld, Randerson, \&
  Falkowski}]{field_primary_1998}
Field, C.~B., Behrenfeld, M.~J., Randerson, J.~T., \& Falkowski, P. 1998,
  Primary production of the biosphere: integrating terrestrial and oceanic
  components, Science, 281, 237.
\newblock \url{http://science.sciencemag.org/content/281/5374/237.short}

\bibitem[{Gausman(1974)}]{gausman_leaf_1974}
Gausman, D.~H. 1974, Leaf {Reflectance} of {Near}-lnfrared, Photogrammetric
  Engineering, 40, 183

\bibitem[{Hill \& Jones(2000)}]{hill_absorption_2000}
Hill, C., \& Jones, R.~L. 2000, Absorption of solar radiation by water vapor in
  clear and cloudy skies: {Implications} for anomalous absorption, Journal of
  Geophysical Research: Atmospheres, 105, 9421, \dodoi{10.1029/1999JD901153}

\bibitem[{Kiang {et~al.}(2007{\natexlab{a}})Kiang, Siefert, {Govindjee}, \&
  Blankenship}]{kiang_spectral_2007-1}
Kiang, N.~Y., Siefert, J., {Govindjee}, \& Blankenship, R.~E.
  2007{\natexlab{a}}, Spectral {Signatures} of {Photosynthesis}. {I}. {Review}
  of {Earth} {Organisms}, Astrobiology, 7, 222, \dodoi{10.1089/ast.2006.0105}

\bibitem[{Kiang {et~al.}(2007{\natexlab{b}})Kiang, Segura, Tinetti,
  {Govindjee}, Blankenship, Cohen, Siefert, Crisp, \&
  Meadows}]{kiang_spectral_2007}
Kiang, N.~Y., Segura, A., Tinetti, G., {et~al.} 2007{\natexlab{b}}, Spectral
  {Signatures} of {Photosynthesis}. {II}. {Coevolution} with {Other} {Stars}
  {And} {The} {Atmosphere} on {Extrasolar} {Worlds}, Astrobiology, 7, 252,
  \dodoi{10.1089/ast.2006.0108}

\bibitem[{Kopparapu {et~al.}(2013)Kopparapu, Ramirez, Kasting, Eymet, Robinson,
  {Suvrath Mahadevan}, Terrien, Domagal-Goldman, Meadows, \&
  Deshpande}]{kopparapu_habitable_2013}
Kopparapu, R.~K., Ramirez, R., Kasting, J.~F., {et~al.} 2013, Habitable {Zones}
  around {Main}-sequence {Stars}: {New} {Estimates}, The Astrophysical Journal,
  765, 131, \dodoi{10.1088/0004-637X/765/2/131}

\bibitem[{Lehmer {et~al.}(2018)Lehmer, Catling, Parenteau, \&
  Hoehler}]{lehmer_productivity_2018}
Lehmer, O.~R., Catling, D.~C., Parenteau, M.~N., \& Hoehler, T.~M. 2018, The
  {Productivity} of {Oxygenic} {Photosynthesis} around {Cool}, {M} {Dwarf}
  {Stars}, The Astrophysical Journal, 859, 171,
  \dodoi{10.3847/1538-4357/aac104}

\bibitem[{Lyons {et~al.}(2014)Lyons, Reinhard, \& Planavsky}]{lyons_rise_2014}
Lyons, T.~W., Reinhard, C.~T., \& Planavsky, N.~J. 2014, The rise of oxygen in
  {Earth}{\textquoteright}s early ocean and atmosphere, Nature, 506, 307,
  \dodoi{10.1038/nature13068}

\bibitem[{Marosv{\"o}lgyi \& van Gorkom(2010)}]{marosvolgyi_cost_2010}
Marosv{\"o}lgyi, M.~A., \& van Gorkom, H.~J. 2010, Cost and color of
  photosynthesis, Photosynthesis Research, 103, 105,
  \dodoi{10.1007/s11120-009-9522-3}

\bibitem[{Mauzerall {et~al.}(1976)Mauzerall, Neuberger, \&
  Kenner}]{mauzerall_chlorophyll_1976}
Mauzerall, D., Neuberger, A., \& Kenner, G.~W. 1976, Chlorophyll and
  photosynthesis, Philosophical Transactions of the Royal Society of London. B,
  Biological Sciences, 273, 287, \dodoi{10.1098/rstb.1976.0014}

\bibitem[{McCree(1971)}]{mccree_action_1971}
McCree, K.~J. 1971, The action spectrum, absorptance and quantum yield of
  photosynthesis in crop plants, Agricultural Meteorology, 9, 191

\bibitem[{Mielke {et~al.}(2011)Mielke, Kiang, Blankenship, Gunner, \&
  Mauzerall}]{mielke_efficiency_2011}
Mielke, S., Kiang, N., Blankenship, R., Gunner, M., \& Mauzerall, D. 2011,
  Efficiency of photosynthesis in a {Chl} d-utilizing cyanobacterium is
  comparable to or higher than that in {Chl} a-utilizing oxygenic species,
  Biochimica et Biophysica Acta (BBA) - Bioenergetics, 1807, 1231,
  \dodoi{10.1016/j.bbabio.2011.06.007}

\bibitem[{Milo(2009)}]{milo_what_2009}
Milo, R. 2009, What governs the reaction center excitation wavelength of
  photosystems {I} and {II}?, Photosynthesis Research, 101, 59,
  \dodoi{10.1007/s11120-009-9465-8}

\bibitem[{O'Malley-James \& Kaltenegger(2018)}]{omalley-james_vegetation_2018}
O'Malley-James, J.~T., \& Kaltenegger, L. 2018, The {Vegetation} {Red} {Edge}
  {Biosignature} {Through} {Time} on {Earth} and {Exoplanets}, Astrobiology,
  18, 1123, \dodoi{10.1089/ast.2017.1798}

\bibitem[{Papageorgiou(2004)}]{papageorgiou_fluorescence_2004}
Papageorgiou, G.~C. 2004, in Chlorophyll a {Fluorescence}: {A} {Signature} of
  {Photosynthesis}, ed. G.~C. Papageorgiou \& {Govindjee}, Advances in
  {Photosynthesis} and {Respiration} (Dordrecht: Springer Netherlands), 43--63,
  \dodoi{10.1007/978-1-4020-3218-9_2}

\bibitem[{Porporato {et~al.}(2002)Porporato, D{\textquoteright}Odorico, Laio,
  Ridolfi, \& Rodriguez-Iturbe}]{porporato_ecohydrology_2002}
Porporato, A., D{\textquoteright}Odorico, P., Laio, F., Ridolfi, L., \&
  Rodriguez-Iturbe, I. 2002, Ecohydrology of water-controlled ecosystems,
  Advances in Water Resources, 25, 1335, \dodoi{10.1016/S0309-1708(02)00058-1}

\bibitem[{Porporato {et~al.}(2003)Porporato, D{\textquoteright}Odorico, Laio,
  \& Rodriguez-Iturbe}]{porporato_hydrologic_2003}
Porporato, A., D{\textquoteright}Odorico, P., Laio, F., \& Rodriguez-Iturbe, I.
  2003, Hydrologic controls on soil carbon and nitrogen cycles. {I}. {Modeling}
  scheme, Advances in Water Resources, 26, 45,
  \dodoi{10.1016/S0309-1708(02)00094-5}

\bibitem[{Reinhard {et~al.}(2016)Reinhard, Planavsky, Gill, Ozaki, Robbins,
  Lyons, Fischer, Wang, Cole, \& Konhauser}]{reinhard_evolution_2016}
Reinhard, C.~T., Planavsky, N.~J., Gill, B.~C., {et~al.} 2016, Evolution of the
  global phosphorus cycle, Nature, 541, 386, \dodoi{10.1038/nature20772}

\bibitem[{Ritchie {et~al.}(2017)Ritchie, Larkum, \& Ribas}]{ritchie_could_2017}
Ritchie, R.~J., Larkum, A.~W., \& Ribas, I. 2017, Could photosynthesis function
  on {Proxima} {Centauri} b?, International Journal of Astrobiology, 17, 147,
  \dodoi{10.1017/S1473550417000167}

\bibitem[{Ross \& Calvin(1967)}]{ross_thermodynamics_1967}
Ross, R.~T., \& Calvin, M. 1967, Thermodynamics of {Light} {Emission} and
  {Free}-{Energy} {Storage} in {Photosynthesis}, Biophysical Journal, 7, 595,
  \dodoi{10.1016/S0006-3495(67)86609-8}

\bibitem[{Sagan {et~al.}(1993)Sagan, Thompson, Carlson, Gurnett, \&
  Hord}]{sagan_search_1993}
Sagan, C., Thompson, W.~R., Carlson, R., Gurnett, D., \& Hord, C. 1993, A
  search for life on {Earth} from the {Galileo} spacecraft, Nature, 365, 715,
  \dodoi{10.1038/365715a0}

\bibitem[{Schwieterman {et~al.}(2018)Schwieterman, Kiang, Parenteau, Harman,
  DasSarma, Fisher, Arney, Hartnett, Reinhard, Olson, Meadows, Cockell, Walker,
  Grenfell, Hegde, Rugheimer, Hu, \& Lyons}]{schwieterman_exoplanet_2018}
Schwieterman, E.~W., Kiang, N.~Y., Parenteau, M.~N., {et~al.} 2018, Exoplanet
  {Biosignatures}: {A} {Review} of {Remotely} {Detectable} {Signs} of {Life},
  Astrobiology, 18, 663, \dodoi{10.1089/ast.2017.1729}

\bibitem[{Seager {et~al.}(2005)Seager, Turner, Schafer, \&
  Ford}]{seager_vegetations_2005}
Seager, S., Turner, E., Schafer, J., \& Ford, E. 2005, Vegetation's {Red}
  {Edge}: {A} {Possible} {Spectroscopic} {Biosignature} of {Extraterrestrial}
  {Plants}, Astrobiology, 5, 372, \dodoi{10.1089/ast.2005.5.372}

\bibitem[{Segura {et~al.}(2005)Segura, Kasting, Meadows, Cohen, Scalo, Crisp,
  Butler, \& Tinetti}]{segura_biosignatures_2005}
Segura, A., Kasting, J.~F., Meadows, V., {et~al.} 2005, Biosignatures from
  {Earth}-{Like} {Planets} {Around} {M} {Dwarfs}, Astrobiology, 5, 706,
  \dodoi{10.1089/ast.2005.5.706}

\bibitem[{Segura {et~al.}(2003)Segura, Krelove, Kasting, Sommerlatt, Meadows,
  Crisp, Cohen, \& Mlawer}]{segura_ozone_2003}
Segura, A., Krelove, K., Kasting, J.~F., {et~al.} 2003, Ozone {Concentrations}
  and {Ultraviolet} {Fluxes} on {Earth}-{Like} {Planets} {Around} {Other}
  {Stars}, Astrobiology, 3, 689, \dodoi{10.1089/153110703322736024}

\bibitem[{Stomp {et~al.}(2007)Stomp, Huisman, Stal, \&
  Matthijs}]{stomp_colorful_2007}
Stomp, M., Huisman, J., Stal, L.~J., \& Matthijs, H. C.~P. 2007, Colorful
  niches of phototrophic microorganisms shaped by vibrations of the water
  molecule, The ISME Journal; London, 1, 271,
  \dodoi{http://dx.doi.org/10.1038/ismej.2007.59}

\bibitem[{Takizawa {et~al.}(2017)Takizawa, Minagawa, Tamura, Kusakabe, \&
  Narita}]{takizawa_red-edge_2017}
Takizawa, K., Minagawa, J., Tamura, M., Kusakabe, N., \& Narita, N. 2017,
  Red-edge position of habitable exoplanets around {M}-dwarfs, Scientific
  Reports, 7, \dodoi{10.1038/s41598-017-07948-5}

\bibitem[{Tinetti {et~al.}(2006)Tinetti, Rashby, \&
  Yung}]{tinetti_detectability_2006}
Tinetti, G., Rashby, S., \& Yung, Y.~L. 2006, Detectability of
  {Red}-{Edge}-shifted {Vegetation} on {Terrestrial} {Planets} {Orbiting} {M}
  {Stars}, The Astrophysical Journal, 644, L129, \dodoi{10.1086/505746}

\bibitem[{Tucker \& Maxwell(1976)}]{tucker_sensor_1976}
Tucker, C.~J., \& Maxwell, E. 1976, Sensor {Design} for {Monitoring}
  {Vegetation} {Canopies}, Photogrammetric Engineering, 42, 1399

\bibitem[{Tyrrell(1999)}]{tyrrell_relative_1999}
Tyrrell, T. 1999, The relative influences of nitrogen and phosphorus on oceanic
  primary production, Nature, 400, 525, \dodoi{10.1038/22941}

\bibitem[{Wang \& He(2021)}]{wang_diurnal_2021}
Wang, F., \& He, J. 2021, Diurnal {Variability} and {Detectability} of
  {Vegetation} {Red} {Edge} of {Earth}-like {Exoplanets}, The Astrophysical
  Journal, 909, 9, \dodoi{10.3847/1538-4357/abd6ff}

\bibitem[{Wolstencroft \& Raven(2002)}]{wolstencroft_photosynthesis:_2002}
Wolstencroft, R., \& Raven, J. 2002, Photosynthesis: {Likelihood} of
  {Occurrence} and {Possibility} of {Detection} on {Earth}-like {Planets},
  Icarus, 157, 535, \dodoi{10.1006/icar.2002.6854}

\end{thebibliography}

\end{document}